\documentclass[subeqn]{article}
  \usepackage[T1]{fontenc}
  \usepackage{amsmath,amssymb,mathrsfs}
  \usepackage[left=4cm, right=4cm, top=3cm, bottom=3cm]{geometry}
  \usepackage{enumitem}
\usepackage[toc,page]{appendix}

\usepackage{hyperref}

\usepackage[ansinew]{inputenc}
\usepackage{cases}
\usepackage{cases}
\usepackage{array}
\usepackage{mathenv}
\usepackage{graphicx,color}
\usepackage{eurosym}
\usepackage{amsmath,amsfonts,amssymb,amsthm}
\usepackage{mathrsfs}
\usepackage{wasysym}
\usepackage{oldstyle}
\usepackage[T1]{fontenc}
\usepackage{pb-diagram}

  \theoremstyle{plain}
   \newtheorem{theorem}{Theorem}
   \newtheorem{lemma}{Lemma}
   \newtheorem{proposition}{Proposition}
  \theoremstyle{definition}
  \newtheorem{defi}{Definition}
  \theoremstyle{remark}
    \newtheorem{remark}{Remark}
   \newtheorem{notation}{Notation}

\title{\textbf{The positronium and the dipositronium in a Hartee-Fock approximation of quantum electrodynamics}}
\author{Sok J\'er\'emy\\
 Ceremade, UMR 7534, Universit\'e Paris-Dauphine,\\
  Place du Mar\'echal de Lattre de Tassigny,\\
  75775 Paris Cedex 16, France.\\ \\
}%

\usepackage{xspace}
\newcommand{\ket}[1]{\ensuremath{|#1\rangle}\xspace}
\newcommand{\bra}[1]{\ensuremath{\langle #1|}\xspace}
\newcommand{\psh}[2]{\ensuremath{\langle #1\,,\,#2\rangle}\xspace}

\newcommand{\ssum}{\ensuremath{\displaystyle\sum}}
\newcommand{\dint}{\ensuremath{\displaystyle\int}}
\newcommand{\diint}{\ensuremath{\displaystyle\iint}}

\newcommand{\g}{\ensuremath{\gamma}}
\newcommand{\G}{\ensuremath{\Gamma}}

\newcommand{\pvac}{\ensuremath{\overline{\boldsymbol{\pi}}}}
\newcommand{\pvt}{\ensuremath{\boldsymbol{\pi}}}

\newcommand{\ns}[2]{\ensuremath{\lVert#2\rVert_{\mathfrak{S}_{#1}}}}
\newcommand{\nlp}[2]{\ensuremath{\lVert#2\rVert_{L^{#1}}}}

\newcommand{\nqq}[1]{\ensuremath{\lVert#1\rVert_{\text{Ex}}}}
\newcommand{\nb}[1]{\ensuremath{\lVert#1\rVert_{\mathcal{B}}}}
\newcommand{\ncc}[1]{\ensuremath{\lVert#1\rVert_{\mathcal{C}}}}

\newcommand{\ov}[1]{\ensuremath{\overline{#1}}}
\newcommand{\un}[1]{\ensuremath{\underline{#1}}}

\newcommand{\llo}{\ensuremath{\log(\Lambda)}}

\newcommand{\ph}{\ensuremath{\varphi}}

\newcommand{\wh}[1]{\ensuremath{\widehat{#1}}}
\newcommand{\la}{\ensuremath{\lambda}}
\newcommand{\La}{\ensuremath{\Lambda}}
\newcommand{\ttr}{\ensuremath{\mathrm{Tr}}}

\newcommand{\wt}[1]{\ensuremath{\widetilde{#1}}}
\newcommand{\D}{\ensuremath{\mathcal{D}^0}}

\newcommand{\hl}{\ensuremath{\mathfrak{H}_\Lambda}}

\newcommand{\PP}{\ensuremath{\mathcal{P}^0_-}}
\newcommand{\PPP}{\ensuremath{\mathcal{P}^0_+}}

\newcommand{\Pup}{\ensuremath{P_{\uparrow}}}
\newcommand{\Pdow}{\ensuremath{P_{\downarrow}}}

\newcommand{\YYY}{\ensuremath{\mathrm{Y}}}
\newcommand{\Cha}{\ensuremath{\mathrm{C}}}
\newcommand{\Isym}{\ensuremath{\mathrm{I}_{\mathrm{s}}}}
\newcommand{\Kbf}{\ensuremath{\mathbf{K}}}
\newcommand{\Lbf}{\ensuremath{\mathbf{L}}}
\newcommand{\Spbf}{\ensuremath{\mathbf{S}}}
\newcommand{\Jbf}{\ensuremath{\mathbf{J}}}
\newcommand{\PPh}{\ensuremath{\Phi_{\mathrm{SU}}}}

\newcommand{\eps}{\ensuremath{\varepsilon}}

\newcommand{\om}{\ensuremath{\omega}}
\newcommand{\ed}[1]{\ensuremath{\widetilde{E}\left(#1\right)}}

\newcommand{\CC}{\ensuremath{\mathbb{C}^4}}
\newcommand{\RR}{\ensuremath{\mathbb{R}^3}}
\begin{document}
\maketitle

\abstract{The Bogoliubov-Dirac-Fock (BDF) model is a no-photon approximation of quantum electrodynamics. It allows to study relativistic electrons in interaction with the Dirac sea. A state is fully characterized by its one-body density matrix, an infinite rank nonnegative projector.

We prove the existence of the para-positronium, the bound state of an electron and  a positron with antiparallel spins, in the BDF model represented by a critical point of the energy functional in the absence of external field. 

We also prove the existence of the dipositronium, a molecule made of two electrons and two positrons that also appears as a critical point. More generally, for any half integer $j\in \tfrac{1}{2}+\mathbb{Z}_+$, we prove the existence of a critical point of the energy functional made of $2j+1$ electrons and $2j+1$ positrons.
}

\tableofcontents

\section{Introduction and main results}

\subsection{The Dirac operator}

Relativistic quantum mechanics is based on the \emph{Dirac operator} $D_0$, which is the Hamiltonian of the free electron. Its expression is \cite{Th}:
\begin{equation}\label{di_dirac_op}
 D_0:=m_ec^2\beta-i\hbar c\ssum_{j=1}^3\alpha_j \partial_{x_j}
\end{equation}
where $m_e$ is the (bare) mass of the electron, $c$ the speed of light and $\hbar$ the reduced Planck constant and $\beta$ and the $\alpha_j$'s are $4\times 4$ matrices defined as follows:
\begin{equation*}\label{di_beta_alpha}
 \beta:=\begin{pmatrix} 
        \mathrm{Id}_{\mathbb{C}^2} & 0\\ 0 & -\mathrm{Id}_{\mathbb{C}^2}
        \end{pmatrix},\ \alpha_j:= \begin{pmatrix} 
         0 & \sigma_j \\ \sigma_j & 0
        \end{pmatrix},\ j\in\{1,2,3\}
\end{equation*}
\begin{equation*}
 \sigma_1:=\begin{pmatrix} 
         0 & 1\\ 1 & 0
        \end{pmatrix},\ \sigma_2:=\begin{pmatrix} 
         0 & -i\\ i & 0
        \end{pmatrix},\ \sigma_3\begin{pmatrix} 
         1 & 0 \\ -1 & 0
        \end{pmatrix}.
\end{equation*}
The operator $D_0$ acts on the Hilbert space $ \mathfrak{H}$:
\begin{equation}\label{di_space_one_electron}
 \mathfrak{H}:=L^2\big(\RR,\CC\big);
\end{equation}
it is self-adjoint on $\mathfrak{H}$ with domain $H^1(\RR,\CC)$. Its spectrum is $\sigma(D_0)=(-\infty,m_ec^2]\cup[m_e c^2,+\infty)$, which leads to the existence of states with arbitrary small energy.

Dirac postulated that all the negative energy states are already occupied by "virtual electrons", with one electron in each state:  by Pauli's principle real electrons can only have a positive energy.

In this interpretation the Dirac sea, composed by those negatively charged virtual electrons, constitutes a polarizable medium that reacts to the presence of an external field. This phenomenon is called the \emph{vacuum polarization}. 

After the transition of an electron of the Dirac sea from a negative energy state to a positive, there is a real electron with positive energy plus the absence of an electron in the Dirac sea. This hole can be interpreted as the addition of a particle with same mass, but opposite charge: the so-called positron.
The existence of this particle was predicted by Dirac in 1931. Although firstly observed in 1929 independently by Skobeltsyn and Chung-Yao Chao, it was recognized in an experiment lead by Anderson in 1932. 

\subsection{Positronium and dipositronium} 

The positronium is the bound state of an electron and a positron. This system was independently predicted by Anderson and Mohorovi$\check{\mathrm{c}}$i\'c in 1932 and 1934 and was experimentally observed for the first time in 1951 by Martin Deutsch. 

It is unstable: depending on the relative spin states of the positron and electron, its average lifetime in vacuum is 125 ps (para-positronium) or 142 ns (ortho-positronium) \cite{karsh}.

Here we are interested in positronium states in the Bogoliubov-Dirac-Fock (BDF) model. 

In a previous paper we have proved the existence of a state that can be interpreted as the ortho-positronium. Our aim in this paper is to find another one that can be interpreted as the para-positronium and to find another state that can be interpreted as the dipositronium, the bound state of two electrons and two positrons.
To find these states, we use symmetric properties of the Dirac operator.

\subsection{Symmetries} 

\noindent -- Following Dirac's ideas, the free vacuum is described by the negative part of the spectrum $\sigma(D_0)$:
\[
P^0_-=\chi_{(-\infty,0)}(D_0).
\]
A correspondence between negative energy states and positron states is given by the \emph{charge conjugation} $\Cha$ \cite{Th}. This is an antiunitary operator that maps $\mathrm{Ran}\,P^0_{-}$ onto $\mathrm{Ran}(1-P^0_{-})$. In our convention \cite{Th} it is defined by the formula:
\begin{equation}\label{di_chargeconj}
\forall\,\psi\in L^2(\RR),\ \Cha\psi(x)=i\beta\alpha_2\overline{\psi}(x),
\end{equation}
where $\overline{\psi}$ denotes the usual complex conjugation. More precisely:
\begin{equation}\label{di_chargeconjprec}
\Cha\cdot \begin{pmatrix}\psi_1\\ \psi_2\\ \psi_2\\\psi_4\end{pmatrix}=\begin{pmatrix}\overline{\psi}_4\\ -\overline{\psi}_3\\ -\overline{\psi}_2\\\overline{\psi}_1\end{pmatrix}.
\end{equation}
In our convention it is also an \emph{involution}: $\Cha^2=\text{id}$. An important property is the following:
\begin{equation}\label{di_denspsi}
\forall\,\psi\in\,L^2,\forall\,x\in\mathbb{R}^3,\ |\Cha \psi(x)|^2=|\psi(x)|^2.
\end{equation}
The Dirac operator anti-commutes with $D_0$, or equivalently there holds
\[
 -\Cha D_0 \Cha^{-1}=-\Cha D_0\Cha=D_0.
\]

\noindent -- There exists another simple symmetry. We define
\begin{equation}\label{di_Isym}
\Isym:=\begin{pmatrix}0 & -\mathrm{Id}_{\mathbb{C}^2}\\
\mathrm{Id}_{\mathbb{C}^2}& 0 \end{pmatrix}\in\mathbb{C}^{4\times 4}.
\end{equation}
This operator is $-i$ the \emph{time reversal operator} $\text{L}_T$ \cite[2.5.7]{Th} in $\mathfrak{H}$, interpreted as a unitary reprsentation of the Poincar{\'e} group.

It acts on the spinor by simple multiplication, furthermore we have $\Isym^2=-\mathrm{Id}$ and
\[
\Isym:\begin{array}{rcl}
\mathrm{Ran}\,P^0_-&\overset{\simeq}{\longrightarrow}& \mathrm{Ran}\,(1-P^0_-)\\
\psi(x)&\mapsto& \Isym\psi(x)
\end{array}
\]
Similarly we have $ -\Isym D_0 \Isym^{-1}=\Isym D_0 \Isym= D_0.$

\noindent -- To end this part we recall that $\mathbf{SU}(2)$ acts on $\mathfrak{H}$ \cite{Th}. Writing $\boldsymbol{\alpha}:=(\alpha_j)_{j=1}^3$ and
\begin{equation}\label{di_L,S}
\mathbf{p}:=-i\hbar\nabla,\ \Lbf:=\mathbf{x}\wedge \mathbf{p},\ \Spbf:=-\frac{i}{4}\boldsymbol{\alpha}\wedge \boldsymbol{\alpha}=\frac{1}{2}\begin{pmatrix}\boldsymbol{\sigma}&0\\ 0&\boldsymbol{\sigma} \end{pmatrix},
\end{equation}
we define
\begin{equation}\label{di_J_moment}
\Jbf:=\Lbf+\Spbf.
\end{equation}
The operator $\mathbf{L}$ is the angular momentum operator and $\mathbf{J}$ is the total angular momentum. From a geometrical point of view, $-i\mathbf{J}$ gives rise to a unitary representation of $\mathbf{SU}(2)$ in $\mathfrak{H}$ by the following formula:
\[
\left\{\begin{array}{l}e^{-i\theta\mathbf{J}\cdot\om}\psi(x)=e^{-i\mathbf{S}\cdot\om}\psi\big( \mathbf{R}^{-1}_{\om,\theta}\big),\\
				\forall\theta\in[0,4\pi),\forall\psi\in\mathfrak{H},\forall\om\in\mathbb{S}^2,
\end{array}\right.
\]
where $\mathbf{R}_{\om,\theta}\in\mathrm{SO}(3)$ is the rotation with axis $\om$ and angle $\theta$. 

As each $\mathrm{S}_j$ is diagonal by block, it is clear that this group representation can be decomposed in two representations, the first acting on the upper spinors $\phi\in L^2(\RR,\mathbb{C}^2)$ and the second on the lower spinors $\chi\in L^2(\RR,\mathbb{C}^2)$:
\[
\psi=:\begin{pmatrix}\phi\\\chi\end{pmatrix}.
\]
In \cite[pp. 122-129]{Th} it is proved that $D_0$ commutes with the action of $\mathbf{SU}(2)$, thus the representation can also be decomposed with respect to $\mathrm{Ran}\,P^0_-$ and $\mathrm{Ran}\,(1-P^0_-)$.

From an algebraic point of view, there exists a group morphism $\PPh:\mathbf{SU}(2)\to \mathbf{U}(\hl)$ where $\mathbf{U}(\mathfrak{H})$ is the set of unitary operator of $\mathfrak{H}$. We write
\begin{equation}
\mathbf{S}:=\PPh\big(\mathbf{SU}(2) \big).
\end{equation}
The irreducible representations of $\PPh$ are known and are expressed in terms of eigenspaces of $\Jbf^2,\Spbf$. The proofs of the following can be found in \cite[pp. 122-129]{Th}.

The operators $\Jbf^2,\mathrm{J}_3,\Kbf$ all commute with each other, and $\Jbf^2,\Kbf$ with $D_0$. Moreover $\Kbf$ commutes with the action $\PPh$.

We have $\hl\subset L^2(\mathbb{R}^3)\simeq L^2((0,\infty),dr)\otimes L^2(\mathbb{S}^2)^4$, and $\Jbf$, $\Lbf$ only act on the part $L^2(\mathbb{S}^2)^4$.

Restricted to $L^2(\mathbb{S}^2)^4$, we have
\begin{equation}
 \sigma\,(\Jbf^2)=\big\{j(j+1),\ j\in\frac{1}{2}+\mathbb{Z}_+\big\},
\end{equation}
and for each eigenvalue $j(j+1)\in \sigma\,\Jbf^2$, the eigenspace $\mathrm{Ker}\big(\Jbf^2-j(j+1)\big)$ may be decomposed with respect to the eigenspaces of $\mathrm{J}_3$ and $\Spbf$. The corresponding eigenvalues are
\begin{enumerate}
 \item $m_j=-j,-j+1,\cdots,j-1,j$ for $\mathrm{J}_3$,
 \item $\kappa_j=\pm\big(j+\frac{1}{2}\big)$ for $\Spbf$.
\end{enumerate}
The eigenspace $\mathfrak{k}_{m_j,\kappa_j}$ of a triplet $(j,m_j,\kappa_j)$ has dimension $2$ and is spanned by $\Phi^+_{m_j,\kappa_j}\perp\Phi^-_{m_j,\kappa_j}$, which have respectively a zero lower spinor and zero upper spinor.

\begin{lemma}\label{di_irreduc}
 For each irreducible subrepresentation $\Phi'_{\mathrm{SU}}$ of $\PPh$, there exists
 \[
  (j,\eps,\mathbf{z}=[z_1:z_2], a_1(r),a_2(r))\in \big(\frac{1}{2}+\mathbb{Z}_+\big)\times\{+,-\}\times\mathbb{C}P^1\times \big(\mathbb{S}L^2((0,\infty),dr)\big)^2,
 \]
such that the representation $\Phi'_{\mathrm{SU}}$ is spanned by $\psi(x)$ defined as follows:
\[
\forall x=r\om\in\RR, \psi(x):=z_1 ra_1(r)\Phi^+_{j, \eps(j+\tfrac{1}{2})}(\om)+z_2 ra_2(r)\Phi^-_{j,\eps(j+\tfrac{1}{2})}.
\]
\end{lemma}
\begin{remark}\label{di_def_mathbb_s}
We recall that for any Hilbert space $\mathfrak{h}$ and any subspace $V\subset \mathfrak{h}$, we define $\mathbb{S}V$ as the unitary vector in $V$:
\[
\mathbb{S}V:=\{x\in V,\ \lVert x\rVert_{\mathfrak{h}}=1\}.
\]
We will use this notation throughout this paper.
\end{remark}

We prove this Lemma in Section \ref{di_proofmanif}. 

\begin{remark}\label{di_rep_type}
 An irreducible subrepresentation of $\PPh$ is characterized by the two numbers $(j,\kappa_j)$. Indeed, the irreducible representations of $\mathbf{SU}(2)$ are known: they can be described by homogeneous polynomials, and for any $n\in \mathbb{Z}_+$, there is but one irreducible representation of dimension $n+1$, up to isomorphism.
 
 In the case of $\PPh$, the two cases $\kappa_j=\pm(j+\tfrac{1}{2})$ are different but \emph{isomorphic}.
\end{remark}
\begin{notation}
 An irreducible subrepresentation of $\PPh$ spanned by an eigenvector of $\Jbf^2$ and $\Kbf$ with respective eigenvalues $j(j+1)$ and $\eps (j+\tfrac{1}{2})$ will be refered as beeing of type $(j,\eps)$ (where $\eps\in\{+,-\}$).
\end{notation}

\begin{notation}
 Throughout this paper we write $\text{Proj}\,E$ to mean the orthonormal projection onto the vector space $E$.
\end{notation}




\subsection{The BDF model}

This model is a no-photon approximation of quantum electrodynamics (QED) which was introduced by Chaix and Iracane in 1989 \cite{CI}, and studied in many papers \cite{stab,ptf,Sc,mf,at,gs,sok}.

It allows to take into account the Dirac vacuum together an electronic system in the presence of an external field.

This is a Hartree-Fock type approximation in which a state of the system "vacuum plus real electrons" is given by an infinite Slater determinant $\psi_1\wedge\psi_2\wedge \cdots$. Such a state is represented by the projector onto the space spanned by the $\psi_j$'s: its so-called one-body density matrix. For instance $P^0_-$ represents the free Dirac vacuum. 

We do not recall the derivation of the BDF model from QED: we refer the reader to \cite{CI,ptf,mf} for full details.

\begin{remark}
To simplify the notations, we choose relativistic units in which, the mass of the electron $m_e$, the speed of light $c$ and $\hbar$ are set to $1$.
\end{remark}
Let us say that there is an external density $\nu$, \emph{e.g.} that of some nucleus. We write $\alpha>0$ the so-called \emph{fine structure constant} (physically $e^2/(4\pi\eps_0\hbar c)$, where $e$ is the elementary charge and $\eps_0$ the permittivity of free space). 

The relative energy of a Hartree-Fock state represented by its 1pdm $P$ with respect to a state of reference ($P^0_-$ in \cite{CI,ptf}) turns out to be a function of $Q=P-P^0_-$, the so-called reduced one-body density matrix. 


A projector $P$ is the one-body density matrix of a Hartree-Fock state in $\mathcal{F}_{\text{elec}}$ \emph{iff} $P-P^0_-$ is Hilbert-Schmidt, that is compact such that its singular values form a sequence in $\ell^2$ \cite[Appendix]{ptf}. 

An ultraviolet cut-off $\Lambda>0$ is needed: we only consider electronic states in
\[
 \hl:=\big\{ f\in\mathfrak{H},\ \text{supp}\,\wh{f}\subset B(0,\La)\big\},
\]
where $\wh{f}$ is the Fourier transform of $f$.

This procedure gives the BDF energy introduced in \cite{CI} and studied in \cite{ptf,Sc}.
\begin{notation}
Our convention for the Fourier transform $\mathscr{F}$ is the following
\[
\forall\,f\in L^1(\RR),\ \wh{f}(p):=\frac{1}{(2\pi)^{3/2}}\dint f(x)e^{-ixp}dx.
\]
\end{notation}
Let us notice that $\hl$ is invariant under $D_0$ and so under $P^0_-$.

We write $\Pi_\La$ for the orthogonal projection onto $\hl$: $\Pi_\La$ is the Fourier multiplier $\mathscr{F}^{-1}\chi_{B(0,\Lambda)}\mathscr{F}$.

By means of a thermodynamical limit, Hainzl \emph{et al.} showed that the formal minimizer and hence the reference state should not be given by $\Pi_\La P^0_-$ but by another projector $\PP$ in $\hl$ that satisfies the self-consistent equation \underline{\emph{in $\hl$}} \cite{mf}:
\begin{equation}\label{di_PP_self}
 \left\{ \begin{array}{ccl}
          \PP-\tfrac{1}{2}&=&-\text{sign}\big(\D\big),\\
          \D&=&D_0\Pi_\La -\dfrac{\alpha}{2}\dfrac{(\PP-\tfrac{1}{2})(x-y)}{|x-y|}
         \end{array}
\right.
\end{equation}
We have $\PP=\chi_{(-\infty,0)}(\D)$. This operator $\D$ was previously introduced by Lieb \emph{et al.} in \cite{ls}. In $\mathfrak{H}$, the operator $\D$ coincides with a bounded, matrix-valued Fourier multiplier whose kernel is $\hl^{\perp}\subset \mathfrak{H}$.

\begin{notation}
Throughout this paper we write 
\begin{equation}
m=\inf \sigma \big(|\D|\big)\ge 1,
\end{equation}
and
\begin{equation}
\PPP:=\Pi_\La-\PP=\chi_{(0,+\infty)}(\D).
\end{equation}
\end{notation}

The resulting BDF energy $\mathcal{E}^\nu_{\text{BDF}}$ is defined on Hartree-Fock states represented by their one-body density matrix $P$:
\[
\mathscr{N}:=\big\{P\in\mathcal{B}(\hl),\ P^*=P^2=P,\ P-\PP\in\mathfrak{S}_2(\hl)\big\}.
\]

We recall that $\mathcal{B}(\hl)$ is the set of bounded operators and that for $p\ge1$, $\mathfrak{S}_p(\hl)$ is the set of compact operators $A$ such that $\ttr\big(|A|^p\big)<+\infty$ \cite{ReedSim,Sim}. In particular $\mathfrak{S}_{\infty}(\hl)$ is the set $\text{Comp}(\hl)$ of compact operators.

This energy depends on three parameters: the fine structure constant $\alpha>0$, the cut-off $\La>0$ and the external density $\nu$. We assume that $\nu$ has finite \emph{Coulomb energy}, that is 
\begin{equation}
\wh{\nu}\ \text{measurable\ and\ }D(\nu,\nu):=4\pi\underset{\RR}{\dint}\frac{|\wh{\nu}(k)|^2}{|k|^2}dk<+\infty.
\end{equation}
The above integral coincides with $\underset{\RR\times\RR}{\iint}\frac{\nu(x)^*\nu(y)}{|x-y|}dxdy$ whenever this last one is well-defined.

\begin{remark}
The same symmetries holds for $\PP$ and $\PPP$: the charge conjugation $\Cha$ and the operator $\Isym$ maps $\mathrm{Ran}\,\PP$ onto $\mathrm{Ran}\,\PPP$. Moreover thanks to \cite[pp. 122-129]{Th} we can easily check that $\D$ also commutes with the action of $\mathbf{SU}(2)$ and with the operators $\Jbf^2$ and $\Kbf$.

\end{remark}

\subsection{Minimizers and critical points}
For $P\in\mathscr{N}$, we have the identity
\begin{equation}\label{di_eqq}
(P-\PP)^2=\PPP (P-\PP)\PPP-\PP(P-\PP)\PP\in \mathfrak{S}_1.
\end{equation}
The charge of a state $P$ is given by the $\PP$-trace of $P-\PP$, defined by the formula:
\begin{align}
\ttr_{\PP}\big(P-\PP\big)&:=\ttr\big(\PP (P-\PP)\PP+\PPP(P-\PP)\PPP\big),\\
                                           &=\text{Dim}\mathrm{Ran}(\PPP)\cap \mathrm{Ran} (P)-\text{Dim}\mathrm{Ran}(\PP)\cap \mathrm{Ran} (1-P).
\end{align}
A minimizer over states with charge $N\in\mathbb{N}$ is interpreted as a ground state of a system with $N$ electrons, in the presence of an external density $\nu$

The existence problem was studied in several papers \cite{at,sok,sokd}: by \cite[Theorem 1]{at}, it is sufficient to check binding inequalities.

The following results hold under technical assumptions on $\alpha$ and $\La$ (different for each result).

In \cite{at}, Hainzl \emph{et al.} proved existence of minimizers for the system of $N$ electrons with $\nu\ge 0$, provided that $N-1<\int \nu$ . 

In \cite{sok}, we proved the existence of a ground state for $N=1$ and $\nu=0$: an electron can bind alone in the vacuum. This surprising result holds due to the vacuum polarization.

In \cite{sokd}, we studied the charge screening effect: due to vacuum polarization, the observed charge of a minimizer $P\neq \PP$ is different from its real charge $\ttr_{\PP}(P-\PP)$. We also proved it is possible to keep track of this effect in the non-relativistic limit $\alpha\to 0$: the resulting limit is an altered Hartree-Fock energy.

Here we are looking for states with an equal number of electrons and positrons, that is we study $\mathcal{E}^0_{\text{BDF}}$ on 
\begin{equation}
\mathscr{M}:=\Big\{P\in\mathscr{N},\ \ttr_{\PP}\big(P-\PP\big)=0\Big\}.
\end{equation}
From a geometrical point of view $\mathscr{M}$ is a Hilbert manifold and $\mathcal{E}^0_{\text{BDF}}$ is a differentiable map on $\mathscr{M}$ (Propositions \ref{di_manim} and \ref{di_gragra}).

We thus seek a critical point on $\mathscr{M}$, that is some $P\in\mathscr{M},\ P\neq\PP$ such that $\nabla \mathcal{E}^0_{\text{BDF}}(P)=0$.

In \cite{pos_sok}, we have found the ortho-positronium by studying the BDF energy restricted to states with the $\Cha$-symmetry:
\begin{equation}\label{di_mm_cc}
P\in\mathscr{M}\text{\ s.t.\ }P+\Cha P\Cha=\mathrm{Id}_{\hl}.
\end{equation}
We write $\mathscr{M}_{\mathscr{C}}$ the set of such states. We will seek the para-positronium in the set $\mathscr{M}_{\mathscr{I}}$ of states having the $\Isym$-symmetry.
\begin{defi}
\begin{equation}\label{di_ii_ss}
\mathscr{M}_{\mathscr{I}}:=\{P\in\mathscr{M}\text{\ s.t.\ }P+\Isym P\Isym^{-1}=P-\Isym P\Isym=\mathrm{Id}_{\hl}\}.
\end{equation}
Equivalently $P\in\mathscr{M}_{\mathscr{I}}$ if and only if $Q:=P-\PP$ is Hilbert-Schmidt and satisfies
\[
-\Isym Q \Isym^{-1}=\Isym Q \Isym=Q.
\]
\end{defi}
We seek a projector $P$ "close" to a state $P_0$ that can be written as:
\begin{equation}\label{di_imagine}
 P_0=\PP+\ket{\Isym\psi_-}\bra{\Isym\psi_-}-\ket{\psi_-}\bra{\psi_-},\ \PPP \psi_-=0.
\end{equation}


To deal with the dipositronium, we impose an additional symmetry: we define $\mathscr{W}\subset \mathscr{M}_{\mathscr{C}}$ as follows.
\begin{defi}
\begin{equation}\label{di_def_w}
\mathscr{W}:=\big\{P\in \mathscr{M}_{\mathscr{C}},\ \forall U\in\mathbf{S},\ UP U^{-1}=P \big\}.
\end{equation}
Equivalently
\[
P\in\mathscr{W}\,\iff\,Q:=P-\PP\mathrm{\ satisfies\ }-\Cha Q\Cha=Q\mathrm{\ and\ }UQU^{-1}=Q,\ \forall\,U\in \mathbf{S}.
\]
\end{defi}

Those sets $\mathscr{M}_{\mathscr{C}},\mathscr{M}_{\mathscr{I}},\mathscr{W}$ have fine properties: they are all submanifolds of $\mathscr{M}$, invariant under the gradient flow of $\mathcal{E}^0_{\text{BDF}}$ (Proposition \ref{di_mani_ci_sym}). 

However while $\mathscr{M}_{\mathscr{C}}$ has two connected components, $\mathscr{M}_{\mathscr{I}}$ has only one connected component and $\mathscr{W}$ has countable connected components. So we may find critical points by searching a minimizer of the BDF energy over the different connected components of $\mathscr{W}$. 
For the para-positronium, a critical point is found by an argument of mountain pass.

\begin{proposition}\label{di_conn_comp}
There is a one-to-one correspondence between the connected components of $\mathscr{W}$ and the set $\mathbb{Z}_2^2[X]$ of polynomials with coefficients in the ring $\mathbb{Z}_2\times \mathbb{Z}_2$. 

Let $P$ be in $\mathscr{W}$. The vector space $E_1:=\mathrm{Ran}\,P\cap \mathrm{Ran}\,\PPP$ has finite dimension and is invariant under $\PPh$. We decompose it into irreducible representations. 

The projector is associated to $\sum_{\ell=1}^{\ell_0}t_\ell X^\ell$ with $t_\ell=(t_{\ell,1};t_{\ell,-1})$ if and only if for any $j\in\tfrac{1}{2}+\mathbb{Z}_+$:
 \begin{enumerate}
  \item The number $b_{j-\tfrac{1}{2},1}$ of irreducible representations of $E_1$ of type $(j,+)$ satisfies $b_{j-\tfrac{1}{2},1}\equiv t_{j-\tfrac{1}{2},1}[2]$.
  \item The number $b_{j-\tfrac{1}{2},-1}$ of irreducible representations of $E_1$ of type $(j,-)$ satisfies $b_{j-\tfrac{1}{2},-1}\equiv t_{j-\tfrac{1}{2},-1}[2]$.
 \end{enumerate}
\end{proposition}

\begin{notation}\label{di_a_c_i}
The symbols $\mathscr{Y}$ and $\YYY$ denotes respectively $\mathscr{C}$ and $\Cha$ or $\mathscr{I}$ and $\Isym$.
 
 Furthermore the different connected components of $\mathscr{W}$ are written $\mathscr{W}_{p(X)}$ with $p(X)\in\mathbb{Z}_2^2[X]$. 
\end{notation}



To state our main Theorems, we need to introduce the mean-field operator.
\begin{notation}[mean-field operator]
An operator $Q\in \mathscr{V}$ is Hilbert-Schmidt and we write $Q(x,y)$ its integral kernel. Its density $\rho_Q$ is defined by the formula 
\begin{equation}\label{di_dens_def}
\forall x\in\RR,\ \rho_Q(x):=\ttr_{\CC}\big(Q(x,x)\big),
\end{equation}
we prove in the next Section that it is well-defined. The mean-field operator $D^{(\La)}_{Q}$ associated to $Q$ \emph{in the vacuum} is :
\begin{equation}\label{di_mean_field}
D^{(\La)}_{Q}:=\Pi_\La\Big(\D +\alpha \big(\rho_Q*\frac{1}{|\cdot|}-\frac{Q(x,y)}{|x-y|}\big)\Big).
\end{equation}
\end{notation}


\begin{theorem}\label{di_main}
 There exist $\alpha_0,L_0,\La_0>0$ such that if
 \[
  \alpha\le \alpha_0;\ \alpha\llo:=L\le L_0\text{\ and\ }\La^{-1}\le \La_0^{-1},
 \]
then there exists a critical point $\ov{P}=\ov{Q}+\PP$ of $\mathcal{E}^0_{\text{BDF}}$ in $\mathscr{M}_{\mathscr{I}}$ that satisfies the following equation.
\begin{equation}
 \exists 0<\mu<m,\ \exists \psi_a\in \mathrm{Ker}\big(D_{\ov{Q}}^{(\La)}-\mu\big),\ \ov{P}=\chi_{(-\infty,0)}\big(D_{\ov{Q}}^{(\La)}\big)+\ket{\psi_a}\bra{\psi_a}-\ket{\Isym \psi_a}\bra{\Isym \psi_a}.
\end{equation}

As $\alpha$ tends to $0$, the upper spinor of $U_{\la}\psi_a:=\la^{3/2}\psi_a(\la(\cdot))$ with $\la:=\tfrac{g'_1(0)^2}{\alpha m}$ tends to a Pekar minimizer.

\noindent -- \emph{We recall that the Pekar energy is defined as follows}
\[
\forall\,\psi\in H^1,\ \mathcal{E}_{\text{PT}}(\psi):=\nlp{2}{\nabla\psi}^2-D\big(|\psi|^2,|\psi|^2\big).
\]
\emph{The infimum over $\mathbb{S}L^2\cap H^1$ is written $E_{\text{PT}}(1)$.}
\end{theorem}

\begin{theorem}\label{di_main_1}
There exist $L_0,\La_0>0$, and for any $j\in\tfrac{1}{2}+\mathbb{Z}_+$, there exists $\alpha_j$ such that if
  \[
  \alpha\le \alpha_j;\ \alpha\llo:=L\le L_0\text{\ and\ }\La^{-1}\le \La_0^{-1},
 \]
 then there exists a minimizer $P_{\mathbf{t}X^{\ell_0}}=Q+\PP$ of $\mathcal{E}^0_{\text{BDF}}$ over the connected component of $\mathscr{W}_{\mathbf{t}X^{\ell_0}}$ with $\mathbf{t}\in\{(1,0),(0,1)\}$.
 
 Moreover there exists $0<\mu_{\ell_0,\mathbf{t}}<1$ and $\psi\in \mathrm{Ker}\big(D_Q^{(\La)}-\mu_{\ell_0,\mathbf{t}}\big)$ such that
 \[
  P_{\mathbf{t}X^{\ell_0}}=\chi_{(-\infty,0)}(D_Q^{(\La)})+\mathrm{Proj}\,\PPh(\psi)-\mathrm{Proj}\,\PPh(\Cha\psi).
 \]
Any upper spinor $\wt{\ph}$ of $\wt{\psi}\in \PPh(\psi)$ can be written as
\[
 \forall\,x=r\om_x\in\RR,\ \wt{\ph}=:ra(r)\sum_{m=-j}^j c_m(\wt{\ph})\Phi^+_{m,\eps(j_0+\tfrac{1}{2})},\ c_m(\wt{\ph})\in\mathbb{C}.
\]

Furthermore, as $\alpha$ tends to $0$, the function $ U_{\la} a(r)=\la^{3/2}a(\la r)$ tends to a minimizer of the energy $\mathcal{E}_{\mathbf{t}X^{\ell_0}}$ over $\mathbb{S}L^2(\mathbb{R}_+,r^2dr)\cap H^1(\mathbb{R}_+,r^2dr):$ 

\begin{equation}\label{di_non_rel_w}
 \mathcal{E}_{\mathbf{t}X^{\ell_0}}\big(f(r)\big):= \ttr\big(-\Delta\, \mathrm{Proj}\,\PPh(rf(r)\Phi^+_{j_0,\eps(\mathbf{t})}) \big)-\nqq{\mathrm{Proj}\,\PPh\,(rf(r)\Phi^+_{j_0,\eps(\mathbf{t})})}^2.
\end{equation}

In particular, the dipositronium corresponds to the case $\ell_0=j_0-\tfrac{1}{2}=0$.
\end{theorem}

\begin{notation}
The minimum is written $E_{\mathbf{t}X^{\ell_0}}^{nr}$ for the non-relativistic energy and $E_{j_0,\eps(\mathbf{t})}$ for the BDF energy over $\mathscr{W}_{\mathbf{t}X^{j_0-1/2}}$.
\end{notation}

\begin{notation}\label{di_eps_t}
For $\mathbf{t}X^{\ell_0}\in \mathbb{Z}_2^2[X]$ as in Theorem \ref{di_main_1}, $\eps(\mathbf{t})\in\{+,-\}$ denotes $+$ if $\mathbf{t}=(1,0)$ or $-$ if $\mathbf{t}=(0,1)$.
\end{notation}

\begin{remark}
 We expect the existence of minimizers over any connected components of $\mathscr{W}$ (associated to $p(X)\in \mathbb{Z}_2^2[X]$), provided that $\alpha$ is smaller than some $\alpha_{p(X)}$.
\end{remark}
\begin{remark}\label{di_prec_non_rel}
 The non-relativistic energy can be computed:
 \begin{equation}
\left\{\begin{array}{rcl}
 \mathcal{E}_{\mathbf{t}X^{\ell_0}}\big(f(r)\big)&:=&(2j_0+1)\underset{0}{\overset{+\infty}{\dint}} \Big[r^2|f'(r)|^2+(j_0+\eps\tfrac{1}{2})(j_0+1+\eps\tfrac{1}{2})|f(r)|^2\Big]dr\\
       && \ \ \ -\underset{\mathbb{R}_+^2}{\diint}r_1^2r_2^2|f(r_1)|^2|f(r_2)|^2w_{j_0,\eps(\mathbf{t})}(r_1,r_2),\\
 w_{j_0,\eps(\mathbf{t})}(r_1,r_2)&:=&\underset{(\mathbb{S}^2)^2}{\diint}\frac{dn_1 dn_2}{|r_1n_1-r_2n_2|}\Big(\ssum_{m_1,m_2}((\Phi^+_{m_1,\eps(j_0+\tfrac{1}{2})})^*\Phi^+_{m_1,\eps(j_0+\tfrac{1}{2})})(n_1) \Big)\\
 &&\ \ \ \times\Big(\ssum_{m_1,m_2}((\Phi^+_{m_1,\eps(j_0+\tfrac{1}{2})})^*\Phi^+_{m_1,\eps(j_0+\tfrac{1}{2})})(n_2) \Big).
\end{array}\right.
\end{equation}

It corresponds to the energy
\[
 \mathcal{E}_{nr}\big(\G\big):= \ttr\big(-\Delta \G\big)-\nqq{\G}^2,\ 0\le \G\le 1,\ \G\in\mathfrak{S}_1(H^1(\RR,\mathbb{C}^2))
\]
restricted to the subspace
\[
\mathscr{S}_{(j_0,\eps(\mathbf{t}))}:=\big\{\G,\ \G^*=\G^2=\G,\ \mathrm{Ran}\,(\PPh)_{\big|_{\G}}\ \text{irreducible\ of\ type\ }(j_0,\eps(\mathbf{t}))\big\}.
\]
This subspace is invariant under the action of $\PPh$ and it is easy to see that it is a submanifold of $ \big\{ \G,\ \G^*=\G^2=\G,\ \ttr\,\G=2j_0+1\big\}$.

The subspace $\mathscr{S}_{(j_0,\eps(\mathbf{t}))}$ is invariant under the flow of $\mathcal{E}_{nr}$.
\end{remark}

The energies can be estimated.
\begin{proposition}\label{di_est}
 In the same regime as in Theorem \ref{di_main}, the following holds. The critical point $\ov{P}$ of the BDF functional over $\mathscr{M}_{\mathscr{I}}$ satisfies
\begin{equation}\label{di_en_para}
 \mathcal{E}^0_{\text{BDF}}(\ov{P})=2m +\frac{\alpha^2m}{g'_1(0)^2}E_{\text{PT}}(1)+\mathcal{O}(\alpha^3).
\end{equation}
Furthermore the minimizer $\ov{P}_{\ell_0}$ over $\mathscr{W}_{\mathbf{t}X^{\ell_0}}$ satisfies:
\begin{equation}\label{di_est_mult}
  \mathcal{E}^0_{\text{BDF}}(\ov{P}_{\ell_0})=2(2j_0+1)+\frac{\alpha^2 m}{g'_1(0)^2} E_{\mathbf{t}X^{\ell_0}}^{nr}+\mathcal{O}(\alpha^3K(j_0)).
\end{equation}
\end{proposition}

\begin{remark}
The Pekar model describes an electron trapped in its own hole in a polarizable medium. Thus it is not surprising to find it here. We recall that there is a unique minimizer of the Pekar energy up to translation and a phase in $\mathbb{S}^7$ (in $\mathbb{C}^4$).

The asymptotic expansion \eqref{di_en_para} coincides with that of the ortho-positronium \cite{pos_sok}. In fact, it can be proved that the first difference between the energies occurs at order $\alpha^4$. 
\end{remark}



\begin{notation}
Throughout this paper we write $K$ to mean a constant independent of $\alpha,\La$. Its value may differ from one line to the other. When we write $K(a)$, we mean a constant that depends solely on $a$. We also use the symbol $\apprle$: $0\le a\apprle b$ means there exists $K>0$ such that $a\le Kb$.

We also recall the reader our use of the notation $\mathbb{S}V$ for any subspace $V$ of some Hilbert space that denotes the set of unitary vector in $V$.
\end{notation}

\subsection{Remarks and notations about $\D$}

$\D$ has the following form \cite{mf}:
\begin{equation}\label{di_D_form}
 \D=g_0(-i\nabla)\beta -i\boldsymbol{\alpha}\cdot \frac{\nabla}{|\nabla|}g_1(-i\nabla)
\end{equation}
where $g_0$ and $g_1$ are smooth radial functions on $B(0,\La)$. Moreover we have:
\begin{equation}
 \forall\,p\in B(0,\La),\ 1\le g_0(p),\text{\ and\ }|p|\le g_1(p)\le |p|g_0(p).
\end{equation}
\begin{notation}
For $\alpha\llo$ sufficiently small, we have $m=g_0(0)$ \cite{LL,sok}.
\end{notation}

\begin{remark}
The smallness of $\alpha$ is needed to get estimates that hold close to the non-relativistic limit.

The smallness of $\alpha\llo$ is needed to get estimates of $\D$: in this case $\D$ can be obtained by a fixed point scheme \cite{mf,LL}, and we have \cite[Appendix A]{sok}:
\begin{equation}\label{di_estim_g}
\begin{array}{c}
g'_0(0)=0,\ \text{and}\ \nlp{\infty}{g'_0},\nlp{\infty}{g_0''}\le K\alpha\\
\nlp{\infty}{g'_1-1}\le K\alpha\llo\le \tfrac{1}{2}\ \text{and}\ \nlp{\infty}{g_1''}\apprle 1.
\end{array}
\end{equation}
\end{remark}


\ \newline


\section{Description of the model}

\subsection{The BDF energy}

\begin{notation}
For any $\eps,\eps'\in\{+,-\}$ and $A\in\mathcal{B}(\hl)$, we write
\begin{equation}
A^{\eps,\eps'}:=\mathcal{P}^0_{\eps}A\mathcal{P}^0_{\eps'}.
\end{equation}
\end{notation}

\begin{notation}
 For an operator $Q\in\mathfrak{S}_2(\hl)$, we write $R_Q$ the operator given by the integral kernel:
 \[
  R_Q(x,y):=\frac{Q(x,y)}{|x-y|}.
 \]
\end{notation}


\begin{defi}[BDF energy]
Let $\alpha>0,\La>0$ and $\nu\in\mathcal{S}'(\RR)$ a generalized function with $D(\nu,\nu)<+\infty$. For $P\in\mathscr{N}$ we write $Q:=P-\PP$ and
\begin{equation}\label{di_formule_bdf}
\left\{\begin{array}{l}
\mathcal{E}^0_{\text{BDF}}(Q)=\ttr_{\PP}\big(\D Q \big)-\alpha D(\rho_Q,\nu)+\dfrac{\alpha}{2}\Big(D(\rho_Q,\rho_Q)-\nqq{Q}^2\Big),\\
\forall\,x,y\in\RR,\ \rho_Q(x):=\ttr_{\mathbb{C}^4}\big(Q(x,x)\big),\ \nqq{Q}^2:=\diint\frac{|Q(x,y)|^2}{|x-y|}dxdy,
\end{array}\right.
\end{equation}
where $Q(x,y)$ is the integral kernel of $Q$. 
\end{defi}
\begin{remark}
The term $\ttr_{\PP}\big(\D Q \big)$ is the kinetic energy, $-\alpha D(\rho_Q,\nu)$ is the interaction energy with $\nu$. The term $\dfrac{\alpha}{2}D(\rho_Q,\rho_Q)$ is the so-called \emph{diract term} and $-\dfrac{\alpha}{2}\nqq{Q}^2$ is the \emph{exchange term}.
\end{remark}

Let us see that formula \eqref{di_formule_bdf} is well-defined whenever $Q$ is $\PP$-trace-class \cite{ptf,at}.

\paragraph{$\mathfrak{S}_1^{\PP}$ and the variational set $\mathcal{K}$} The set $\mathfrak{S}_1^{\PP}$ of $\PP$-trace class operator is the following Banach space:
\begin{equation}
\mathfrak{S}_1^{\PP}=\big\{Q\in\mathfrak{S}_2(\hl),\ Q^{++},Q^{--}\in\mathfrak{S}_1(\hl)\big\},
\end{equation} 
with the norm
\begin{equation}
\lVert Q\rVert_{\mathfrak{S}_1^{\PP}}:=\ns{2}{Q^{+-}}+\ns{2}{Q^{-+}}+\ns{1}{Q^{++}}+\ns{1}{Q^{--}}.
\end{equation}


We have $\mathscr{N}\subset \PP+\mathfrak{S}_1^{\PP}$ thanks to \eqref{di_eqq}. The closed convex hull of $\mathscr{N}-\PP$ under $\mathfrak{S}_1^{\PP}$ is
\[
\mathcal{K}:=\big\{Q\in\mathfrak{S}_1^{\PP}(\hl),\ Q^*=Q,\ -\PP\le Q\le \PPP\big\}
\]
and we have \cite{ptf,Sc}
\[
\forall\,Q\in \mathcal{K},\ Q^2\le Q^{++}-Q^{--}.
\]

\paragraph{The BDF energy for $Q\in \mathfrak{S}_1^{\PP}$} 
We have
\[
\PP(\D Q)\PP=-|\D|Q^{--}\in\,\mathfrak{S}_1(\hl),\ \text{because}\, |\D|\in\mathcal{B}(\hl),
\]
this proves that the kinetic energy is defined.

By the Kato-Seiler-Simon (KSS) inequality \cite{Sim}, $Q$ is locally trace-class:
\[
 \forall\,\phi\in \mathbf{C}^\infty_0(\RR),\ \phi \Pi_\La\in\mathfrak{S}_2\text{\ so\ }\phi Q \phi=\phi\Pi_\La Q\phi\in\mathfrak{S}_1(L^2(\RR)).
\]
We recall this inequality states that for all $2\le p\le \infty$ and $d\in\mathbb{N}$, we have
\[
\forall\,f,g\in L^p(\mathbb{R}^d),\ f(x)g(-i\nabla)\in\mathfrak{S}_{p}(\hl)\text{\ and\ }\ns{p}{f(x)g(-i\nabla)}\le (2\pi)^{-d/p}\nlp{p}{f}\nlp{p}{g}.
\]
It follows that the \emph{density} $\rho_Q$ of $Q$, defined in \eqref{di_formule_bdf} is well-defined. By the KSS inequality, we can also prove that $\ncc{\rho_Q}\apprle K(\La)\lVert Q \rVert_{\mathfrak{S}_1^{\PP}}$ \cite[Proposition 2]{gs}.

By Kato's inequality:
\begin{equation}\label{di_kato}
\dfrac{1}{|\cdot|}\le \dfrac{\pi}{2}|\nabla|,
\end{equation}
the exchange term is well-defined.

Moreover the following holds: if $\alpha < \tfrac{4}{\pi}$, then the BDF energy is bounded from below on $\mathcal{K}$ \cite{stab,Sc,at}. We have
\begin{equation}\label{di_below}
\forall\,Q_0\in\mathfrak{S}_2(\hl),\ \mathcal{E}^0_{\text{BDF}}(Q_0)\ge \big(1-\alpha\frac{\pi}{4}\big)\ttr\big(|\D||Q_0|^2\big).
\end{equation}

Here we assume it is the case. This result will be often used throughout this paper.


\paragraph{Minimizers} For $Q\in\mathcal{K}$, its charge is its $\PP$-trace: $q=\ttr_{\PP}(Q)$. We define the Charge sector sets:
\[
\forall\,q\in\RR,\ \mathcal{K}^q:=\big\{Q\in\mathcal{K},\ \ttr(Q)=q\big\}.
\]
A minimizer of $\mathcal{E}^\nu_{\text{BDF}}$ over $\mathcal{K}$ is interpreted as the polarized vacuum in the presence of $\nu$ while a minimizer over charge sector $N\in\mathbb{N}$ is interpreted as the ground state of $N$ electrons in the presence of $\nu$, by Lieb's principle \cite[Proposition 3]{at}, such a minimizer is in $\mathscr{N}-\PP$.

We define the energy functional $E^\nu_{\text{BDF}}$:
\begin{equation}
 \forall\,q\in\RR,\ E^\nu_{\text{BDF}}(q):=\inf\big\{\mathcal{E}^\nu_{\text{BDF}}(Q),\ Q\in\mathcal{K}^q\big\}.
\end{equation}

We also write:
\begin{equation}\label{di_koci}
\mathcal{K}^0_{\mathscr{Y}}:=\{ Q\in\mathcal{K},\ \text{Tr}_{\PP}(Q)=0,\ -\YYY Q \YYY^{-1} =Q\}.
\end{equation}
Proposition \ref{di_weakclosed} states that this set is sequentially weakly-$*$ closed in $\mathfrak{S}_1^{\PP}(\hl)$.

\subsection{Structure of manifold}

We consider
\[
 \mathscr{V}=\big\{P-\PP,\ P^*=P^2=P\in\mathcal{B}(\hl),\ \ttr_{\PP}\big( P-\PP\big)=0\big\}\subset \mathfrak{S}_2(\hl). 
\]
and write: $\mathscr{M}:=\PP+\mathscr{V}=\big\{P,\ P^*=P^2=P,\ \ttr_{\PP}\big( P-\PP\big)=0\big\}.$

\medskip

We recall the following proposition, proved in \cite{pos_sok}.
\begin{proposition}\label{di_manim}
The set $\mathscr{M}$ is a Hilbert manifold and for all $P\in\mathscr{M}$,
\begin{equation}
 \mathrm{T}_P \mathscr{M}=\{ [A,P],\,A\in\mathcal{B}(\hl),\ A^*=-A\text{\ and\ }PA(1-P)\in\mathfrak{S}_2(\hl)\}.
\end{equation}
Writing
\begin{equation}
 \mathfrak{m}_P:=\{ A\in\mathcal{B}(\hl),\ A^*=-A,\ PAP=(1-P)A(1-P)=0\text{\ and\ }PA(1-P)\in\mathfrak{S}_2(\hl)\},
\end{equation}
any $P_1\in\mathscr{M}$ can be written as $P_1=e^A P e^{-A}$ where $A\in\mathfrak{m}_P$.
\end{proposition}

The BDF energy $\mathcal{E}_{\text{BDF}}^\nu$ is a differentiable function in $\mathfrak{S}_1^{\PP}(\hl)$ with: 
\begin{equation}\label{di_eqdebdf}
 \left\{ \begin{array}{l}
        \forall\, Q,\delta Q\in\mathfrak{S}_1^{\PP}(\hl),\ \text{d}\mathcal{E}_{\text{BDF}}^\nu(Q)\cdot \delta Q=\text{Tr}_{\PP}\big(D_{Q,\nu}\delta Q\big).\\
        D_{Q,\nu}:=\D+\alpha \big((\rho_Q-\nu)*\frac{1}{|\cdot|}-R_Q\big).
\end{array}\right.
\end{equation}
We may rewrite \eqref{di_eqdebdf} as follows:
\begin{equation}
\forall\, Q,\delta Q\in\mathfrak{S}_1^{\PP}(\hl),\ \text{d}\mathcal{E}_{\text{BDF}}^\nu(Q)\cdot \delta Q=\text{Tr}_{\PP}\big(\Pi_\Lambda D_{Q,\nu}\Pi_\Lambda \delta Q\big)
\end{equation}
We recall the mean-field operator $ D_Q^{(\La)}$ is defined in Notation \ref{di_mean_field}.

\begin{proposition}\label{di_gragra}
Let $(P,v)$ be in the tangent bundle $\mathrm{T}\mathscr{M}$ and $Q=P-\PP$. Then we have $[[\Pi_\Lambda D_Q \Pi_\Lambda,P],P]\in\mathrm{T}_P\mathscr{M}$ and:
 \begin{equation}\label{di_difftan}
\mathrm{d}\mathcal{E}_{\text{BDF}}^0(P)\cdot v=\text{Tr}\Big(\big[\big[ D_{Q}^{(\La)} ,P\big],P\big]v\Big).
\end{equation}
In other words: 
 \begin{equation}\label{di_defgradient}
  \forall\,P\in\mathscr{M},\ \nabla \mathcal{E}_{\text{BDF}}^0(P)=\big[\big[\Pi_\La D_Q \Pi_\La,P\big],P\big].
 \end{equation}
\end{proposition}
\begin{remark}
 The operator $[[\Pi_\La D_Q\Pi_\La,P],P]$ is the "projection" of $\Pi_\La D_Q \Pi_\La$ onto $\text{T}_P\mathscr{M}$. 
\end{remark}

In \cite{pos_sok}, we proved that $\mathscr{M}_{\mathscr{C}}$ is a submanifold of $\mathscr{M}$. We recall that the notations $\mathscr{Y}$, $\YYY$ are specified in Notation \ref{di_a_c_i}.

\begin{proposition}\label{di_mani_ci_sym}
The sets $\mathscr{M}_{\mathscr{I}}$ and $\mathscr{W}$ are \emph{submanifolds} of $\mathscr{M}$, which are \emph{invariant} under the flow of $\mathcal{E}_{\text{BDF}}^0$. The following holds: for any $P\in \mathscr{M}_{\mathscr{Y}}$, writing
\begin{equation}
 \mathfrak{m}^{\mathscr{Y}}_P=\{a\in \mathfrak{m}_P,\ \YYY a \YYY^{-1}=a\},
\end{equation}
we have
\begin{equation}\label{di_tangentc}
 \mathrm{T}_P \mathscr{M}_{\mathscr{Y}}=\{[a,P],\ a\in \mathfrak{m}_P^{\mathscr{Y}}\}=\{v\in\mathrm{T}_P \mathscr{M},\ -\YYY v \YYY^{-1}=v\}.
\end{equation}
Furthermore, for any $P\in\mathscr{M}_{\mathscr{Y}}$ we have $\rho_{P-\PP}=0.$

For $P\in\mathscr{W}$, the same holds with
\[
\left\{ \begin{array}{rcl}
 \mathfrak{m}^{\mathscr{W}}_{P}&:=&\big\{a\in \mathfrak{m}^{\mathscr{C}}_{P},\ \forall\,U\in\mathbf{S},\ U a U^{-1}=a\big\},\\
 \mathrm{T}_P\mathscr{W}&:=&\big\{[a,P],\ a\in\mathfrak{m}_P^{\mathscr{W}}\big\}.
	\end{array}
\right.
\]
\end{proposition}

\begin{remark}[Lagrangians]\label{Lagrangians}
The operator $\Isym$ induced a symplectic structure on the \emph{real} Hilbert space $(\hl,\mathfrak{Re}\psh{\cdot}{\cdot}_{\mathfrak{H}})$:
\[
\forall\,f,g\in\hl,\ \om_{\mathrm{I}}(f,g):=\mathfrak{Re}\psh{f}{\Isym g}.
\]
The manifold $\mathscr{M}_{\mathscr{I}}$ is constituted by \emph{Lagrangians} of  $\om_{\mathrm{I}}$ that are in $\mathscr{M}$.
\end{remark}

We end this section by stating technical results.

\subsection{Form of trial states}

The following Theorem is stated in \cite[Appendix]{at} and proved in \cite{pos_sok}.

\begin{theorem}[Form of trial states]\label{di_structure}
Let $P_1,P_0$ be in $\mathscr{N}$ and $Q=P_1-P_0$. Then there exist $M_+,M_-\in\mathbb{Z}_+$ such that there exist two orthonormal families 
\[
\begin{array}{ll}
(a_1,\ldots,a_{M_+})\cup(e_i)_{i\in\mathbb{N}}& \mathrm{in}\ \mathrm{Ran}\,\PPP,\\
(a_{-1},\ldots,a_{-M_+})\cup(e_{-i})_{i\in\mathbb{N}}&\mathrm{in}\ \mathrm{Ran}\,\PP,
\end{array}
\]
and a nonincreasing sequence $(\la_i)_{i\in\mathbb{N}}\in\ell^2$ satisfying the following properties:
\begin{enumerate}
\item The $a_i$'s are eigenvectors for $Q$ with eigenvalue $1$ (resp. $-1$) if $i>0$ (resp. $i<0$).
\item For each $i\in\mathbb{N}$ the plane $\Pi_i:=\text{Span}(e_i,e_{-i})$ is spanned by two eigenvectors $f_i$ and $f_{-i}$ for $Q$ with eigenvalues $\la_i$ and $-\la_i$.
\item The plane $\Pi_i$ is also spanned by two orthogonal vectors $v_i$ in $\mathrm{Ran}(1-P)$ and $v_{-i}$ in $\mathrm{Ran}(P)$. Moreover $\la_i=\sin(\theta_i)$ where $\theta_i\in (0,\tfrac{\pi}{2})$ is the angle between the two lines $\mathbb{C}v_i$ and $\mathbb{C}e_i$.
\item There holds: \[
Q=\ssum_i^{M_+}\ket{a_i}\bra{a_i}-\ssum_i^{M_-}\ket{a_{-i}}\bra{a_{-i}}+\ssum_{j\in \mathbb{N}}\la_j(\ket{f_j}\bra{f_j}-\ket{f_{-j}}\bra{f_{-j}}).
\]
\end{enumerate}
\end{theorem}

\begin{remark}
We have
 \begin{equation}\label{di_++--}
 \begin{array}{l}
  Q^{++}=\ssum_i^{M_+}\ket{a_i}\bra{a_i}+\ssum_{j\in\mathbb{N}}\sin(\theta_j)^2\ket{e_j}\bra{e_j},\\
  Q^{--}=-\ssum_i^{M_-}\ket{a_{-i}}\bra{a_{-i}}-\ssum_{j\in\mathbb{N}}\sin(\theta_j)^2\ket{e_{-j}}\bra{e_{-j}}.
 \end{array}
\end{equation}
\end{remark}

Thanks to Theorem \ref{di_structure}, it is possible to characterize states in $\mathscr{M}_{\mathscr{Y}}$ and $\mathscr{W}$. We restate a proposition of \cite{pos_sok} and add the case of $\Isym$.

\begin{proposition}\label{di_chasym}
Let $\g=P-\PP$ be in $\mathscr{M}_{\mathscr{Y}}$. For $-1\le \mu\le 1$ and $X\in\{\g,\g^2\}$, we write 
\[E^X_\mu=\mathrm{Ker}(X-\mu).\]
Then for any $\mu\in\sigma(\g)$, $\YYY E^\g_\mu=E^\g_{-\mu}$. Moreover for $|\mu|<1$ if we decompose $E^\g_{\mu}\oplus E^\g_{-\mu}$ into a sum of planes $\Pi$ as in Theorem \ref{di_structure}, then 
\begin{enumerate}
 \item If $\YYY=\Isym$, then we can choose the $\Pi$'s to be $\Isym$-invariant.
 \item If $\YYY=\Cha$, then each $\Pi$ is \emph{not} $\Cha$-invariant and $\mathrm{Dim}\,E^\g_{\mu}$ is even. 
 
 Equivalently $\text{Dim}\,E^{\g^2}_{\mu^2}$ is divisible by $4$. Moreover there exists a decomposition 
\[
E^{\g^2}_{\mu^2}=\underset{1\le j\le \tfrac{N}{2}}{\overset{\perp}{\oplus}}V_{\mu,j}\text{\ and\ }V_{\mu,j}=\Pi^a_{\mu,j}\overset{\perp}{\oplus}\Cha \Pi^a_{\mu,j}
\]
where the $\Pi^a_{\mu,j}$'s and $\Cha \Pi^a_{\mu,j}$'s are spectral planes described in Theorem \ref{di_structure}.
\end{enumerate}

\end{proposition}






\subsection{The Cauchy expansion}

In this part, we introduce a useful trick in the model. The Cauchy expansion \eqref{di_cauchy20} is an application of functional calculus: we refer the reader to \cite{ptf,sok} for further details.

We assume $Q_0\in \mathfrak{S}_2$ with 
\begin{equation}\label{di_supp}
\alpha \ns{2}{|\D|^{1/2}Q_0}\ll 1.
\end{equation}

We recall the following inequality, proved in \cite{sok}
\begin{equation}\label{di_cauchy_est0}
 \forall\,Q_0\in\mathfrak{S}_2,\ \ns{2}{R_{Q_0}\tfrac{1}{|\nabla|^{1/2}}}^2\apprle \nqq{Q}^2\apprle \diint |p+q||\wh{Q}(p,q)|^2dpdq,
\end{equation}

From now on, we only deal with $Q_0$ whose density vanishes: $\rho_{Q_0}=0$.
The mean-field operator $D_{Q_0}^{(\La)}$ is away from $0$ thanks to \eqref{di_supp}. Indeed, there holds

\begin{align*}
 |\Pi_\La R_{Q_0}\Pi_\La|^2&\le |\nabla|^{1/2}\,\frac{\Pi_\La}{|\nabla|^{1/2}}R_{Q_0}^* R_{Q_0}\frac{\Pi_\La}{|\nabla|^{1/2}}\,|\nabla|^{1/2}\\
          &\le \Pi_\La|\nabla|\nb{\tfrac{1}{|\nabla|^{1/2}} R_{Q_0}}^2\\
          &\apprle \Pi_\La|\nabla|\nqq{Q_0}\apprle |\D|^2\nqq{Q_0}^2,
\end{align*}
thus
\begin{equation}
|D_{Q_0}^{(\La)}|\apprge |\D|\big(1-\alpha K\nqq{Q_0}\big).
\end{equation}

The Cauchy expansion gives an expression of
\[
\g_0:=\chi_{(-\infty,0)}\big(D_{Q_0}^{(\La)}\big)-\PP:=\pvac_0.
\]

We have \cite{ptf}
\begin{equation}\label{di_cauchy0}
\chi_{(-\infty,0)}\big(D_{Q_0}^{(\La)}\big)-\PP=\frac{1}{2\pi}\dint_{-\infty}^{+\infty}\frac{d \om}{\D+i\omega}\big(\alpha\Pi_\La R_{Q_n}\Pi_\La \big)\dfrac{1}{D_{Q_0}+i\om}\Pi_\La.
\end{equation}

We also expand in power of $Y[Q_0]:=-\alpha \Pi_\La R_{Q_0}\Pi_\La $:
\begin{equation}\label{di_cauchy20}
\left\{
	\begin{array}{rcl}
		\pvt_-^n-\PP&=&\ssum_{j\ge 1}\alpha^j M_j[Y[Q_0]],\\
		M_j[Y_n]&=&-\dfrac{1}{2\pi}\dint_{-\infty}^{+\infty}\frac{ d\om}{\D+i\om}\Big(Y_n\frac{1}{\D+i\om} \Big)^{j}.
	\end{array}
\right.
\end{equation}
Each $M_j[Y[Q_0]]$ is polynomial in $\Pi_\La R_{Q_0} \Pi_\La$ of degree $j$. 

By using \eqref{di_cauchy_est0}, the decomposition \eqref{di_cauchy20} is well-defined in several Banach space, provided that $\alpha \nqq{Q_0}$ is small enough.

\noindent -- First, integrating the norm of bounded operator in \eqref{di_cauchy0}, we obtain
\[
\nb{\pvac_0-\PP}\apprle \alpha \nqq{Q_0}<1.
\]

\noindent -- We take the Hilbert-Schmidt norm \cite{ptf,sok}: we get
\begin{equation}\label{di_estim_gn}
\ns{2}{\g_{0}}\apprle \alpha \nqq{Q_0}.
\end{equation}
\noindent -- We take the norm $\ns{2}{|\D|^{1/2}(\cdot)}$ we get the rough estimate
\begin{equation}\label{di_estim_kin}
\ns{2}{|\D|^{1/2}\g_{0}}\apprle \min(\sqrt{L\alpha}\nqq{Q_0},\alpha \ns{2}{R_{Q_0}}\big)+\alpha^2\nqq{Q_0}^2.
\end{equation}

\begin{remark}\label{di_diff_ch}
The same estimates holds for the differential of $Q_0\mapsto \g_0$, for sufficiently small $\alpha$. As shown in \cite{sok}, the upper bound of each norm is a power series of kind
\[
\lVert\g_0\rVert\le \alpha \lVert M_1[Y[Q_0]]\rVert+\ssum_{j=1}^{+\infty}\sqrt{j}\alpha^j\big(K\nqq{ Q_0} \big)^j.
\]
In the case of the differential, we get an upper bound of kind
\[
\lVert \text{d}\g_0\rVert\le \alpha \lVert M_1[Y[Q_0]]\rVert+\ssum_{j=1}^{+\infty}j^{3/2}\alpha^j\big(K\nqq{ Q_0} \big)^j.
\]
The power series converge for sufficiently small $\alpha \nqq{Q_0}$.
\end{remark}

\noindent -- It is also possible to consider other norms, using from the fact that a (scalar) Fourier multiplier $F(\mathbf{p}-\mathbf{q})=F(-i\nabla_x+i\nabla_y)$ commutes with the operator $R[\cdot]:Q(x,y)\mapsto \tfrac{Q(x,y)}{|x-y|}$. We can also consider the norm
\[
\lVert Q_0\rVert_{w}^2:=\diint w(p-q)(\ed{p}+\ed{q})|\wh{Q}_0(p,q)|^2dpdq,
\]
where $w(\cdot)\ge 0$ is any weight satisfying a subadditive condition \cite{sok}:
\[
\forall\,p,q\in\RR,\ \sqrt{w(p+q)}\le K(w)\big(\sqrt{w(p)}+\sqrt{w(q)}\big).
\]

\section{Proof of Theorems \ref{di_main} and \ref{di_main_1}}
\subsection{Strategy and tools of the proof: the dipositronium}
\subsubsection{Topologies}



The existence of a minimizer over $\mathscr{W}_{\mathbf{t}X^\ell}$ (with $\mathbf{t}\in\mathbb{Z}_2^2$) is proved with the same method used in \cite{pos_sok}.

We use a lemma of Borwein and Preiss \cite{borw,at}, a smooth generalization of Ekeland's Lemma \cite{ek}: we study the behaviour of a specific minimizing sequence $(P_n)_n$ or equivalently $(P_n-\PP=:Q_n)_n$. 

This sequence satisfies an equation close to the one satisfied by a real minimizer and we show this equation remains in some weak limit.



\begin{remark}\label{di_topo}
 We recall different topologies over bounded operator, besides the norm topology $\nb{\cdot}$ \cite{ReedSim}.
 \begin{enumerate}
  \item The so-called \emph{strong topology}, the weakest topology $\mathcal{T}_s$ such that for any $f\in\hl$, the map
  \[
   \begin{array}{rcl}
    \mathcal{B}(\hl)&\longrightarrow&\hl\\
    A&\mapsto& Af
   \end{array}
  \]
is continuous.
\item The so-called \emph{weak operator topology}, the weakest topology $\mathcal{T}_{w.o.}$ such that for any $f,g\in\hl$, the map
  \[
   \begin{array}{rcl}
    \mathcal{B}(\hl)&\longrightarrow&\mathbb{C}\\
    A&\mapsto& \psh{A f}{g}
   \end{array}
  \]
  is continuous.
 \end{enumerate}
 We can also endow $\mathfrak{S}_1^{\PP}$ with its weak-$*$ topology, the weakest topology such that the following maps are continuous:
 \[
   \begin{array}{|l}
   \begin{array}{rcl}
    \mathfrak{S}_1^{\PP}&\longrightarrow&\mathbb{C}\\
    Q&\mapsto& \ttr\big(A_0(Q^{++}+Q^{--})+A_2(Q^{+-}+Q^{-+})\big)
   \end{array}\\
  \forall\,(A_0,A_2)\in\mathrm{Comp}(\hl)\times \mathfrak{S}_2(\hl).
  \end{array}
 \]
\end{remark}

\begin{lemma}\label{di_weakclosed}
 The set $\mathcal{K}^0_{\mathscr{Y}}$, defined in \eqref{di_koci}, is weakly-$*$ sequentially closed in $\mathfrak{S}_1^{\PP}(\hl)$.
 
\end{lemma}
\begin{remark}
 This Lemma was stated for $\mathscr{Y}=\mathscr{C}$ in \cite{pos_sok}. For $\mathscr{Y}=\mathscr{I}$ the proof is the same and we refer the reader to this paper.
\end{remark}

\subsubsection{The Borwein and Preiss Lemma}
We recall this Theorem as stated in \cite{at}:

\begin{theorem}\label{di_bp_lemma}
 Let $\mathcal{M}$ be a closed subset of a Hilbert space $\mathcal{H}$, and $F:\mathcal{M}\to (-\infty,+\infty]$ be a lower semi-continuous function that is bounded from below and not identical to $+\infty$.
 For all $\eps>0$ and all $u\in \mathcal{M}$ such that $F(u)<\inf_{\mathcal{M}}+\eps^2$, there exist $v\in\mathcal{M}$ and $w\in\ov{\mathrm{Conv}(\mathcal{M})}$ such that
 
 \begin{enumerate}
  \item $F(v)< \inf_{\mathcal{M}}+\eps^2$,
  \item $\lVert u-v\rVert_{\mathcal{H}}<\sqrt{\eps}$ and $\lVert v-w\rVert_{\mathcal{H}}<\sqrt{\eps}$,
  \item $F(v)+\eps \lVert v-w\rVert_{\mathcal{H}}^2=\min\big\{F(z)+\eps \lVert z-w\rVert_{\mathcal{H}}^2,\ z\in\mathcal{M}\big\}.$
 \end{enumerate}
\end{theorem}

\medskip

\noindent -- Here we apply this Theorem with $\mathcal{H}=\mathfrak{S}_2(\hl)$, $\mathcal{M}=\mathscr{W}_{p(X)}-\PP$ and $F=\mathcal{E}^0_{\mathrm{BDF}}$.

The BDF energy is continuous in the $\mathfrak{S}_1^{\PP}$-norm topology, thus its restriction over $\mathscr{V}$ is continuous in the $\mathfrak{S}_2(\hl)$-norm topology.

This subspace $\mathcal{H}$ is closed in the Hilbert-Schmidt norm topology because $\mathscr{V}=\mathscr{M}_{\mathscr{C}}$ is closed in $\mathfrak{S}_2(\hl)$ and $\mathscr{E}_{-1}-\PP$ is closed in $\mathscr{V}$.

Moreover, we have
\[
 \ov{\text{Conv}(\mathscr{W}_{p(X)}-\PP)}^{\mathfrak{S}_2}\subset \mathcal{K}_{\mathscr{C}}^0.
\]

\medskip

\noindent -- For every $\eta>0$, we get a projector $P_\eta\in\mathscr{W}_{p(X)}$ and $A_\eta\in \mathcal{K}_{\mathscr{C}}^0$ such that $P$ that minimizes the functional $F_\eta: P\in\mathscr{E}_{-1}\mapsto \mathcal{E}_{\text{BDF}}^0(P-\PP)+\eps \ns{2}{P-\PP-A_\eta}^2.$

We write
\begin{equation}\label{di_almost}
 Q_\eta:= P_\eta -\PP,\ \G_\eta:=Q_\eta -A_\eta,\ \wt{D}_{Q_\eta}:=\Pi_\La \big(\D-\alpha R_{Q_\eta}+2\eta \G_\eta\big)\Pi_\La.
\end{equation}
Studying its differential on $\text{T}_{P_\eta} \mathscr{W}$, we get:
\begin{equation}\label{di_eq_almost}
 \big[\wt{D}_{Q_\eta}, P_\eta\big]=0.
\end{equation}
In particular, by functional calculus, we have:
\begin{equation}\label{di_pimoins}
 \big[\boldsymbol{\pi}_-^{\eta},P_\eta\big]=0,\ \boldsymbol{\pi}_{\eta}^-:=\chi_{(-\infty,0)}(\wt{D}_{Q_\eta}).
\end{equation}
We also write
\begin{equation}\label{di_piplus}
 \boldsymbol{\pi}_{\eta}^+:=\chi_{(0,+\infty)}(\wt{D}_{Q_\eta})=\Pi_\La-\boldsymbol{\pi}_{\eta}^-.
\end{equation}
We decompose $\hl$ as follows (here R means $\mathrm{Ran}$):
\begin{equation}\label{di_decomp_hl}
 \hl=\text{R}(P_\eta)\cap \text{R}(\pvt_{\eta}^-)\overset{\perp}{\oplus}\text{R}(P_\eta)\cap \text{R}(\pvt_{\eta}^+)\overset{\perp}{\oplus}\text{R}(\Pi_\La-P_\eta)\cap \text{R}(\pvt_{\eta}^-)\overset{\perp}{\oplus}\text{R}(\Pi_\La-P_\eta)\cap \text{R}(\pvt_{\eta}^+).
\end{equation}

We will prove
\begin{enumerate}
 \item $\mathrm{Ran}\,P\cap \mathrm{Ran}\,\pvt_{\eta}^+$ has dimension $2j+1$ and is invariant under $\PPh$, spanned by a unitary $\psi_\eta\in\hl$.
 \item As $\eta$ tends to $0$, up to translation and a subsequence, $\psi_\eta\rightharpoonup \psi_a\neq 0$, $Q_\eta\rightharpoonup \ov{Q}$.
       There holds $\ov{P}_{j_0}=\ov{Q}+\PP\in \mathscr{W}_{p(X)}$, $\psi_a$ is a unitary eigenvector of $ D_{\ov{Q}}^{(\La)} $ and
       \begin{equation}\label{di_eq_min}
        \ov{Q}+\PP=\chi_{(-\infty,0)}\big( D_{\ov{Q}}^{(\La)}  \big)+\text{Proj}\,\PPh(\psi_a)-\text{Proj}\,\PPh(\Cha \psi_a),
       \end{equation}
       where $\text{Proj}\,E$ means the orthonormal projection onto the vector space $E$.
\end{enumerate}

In the following part we write the spectral decomposition of trial states and prove Lemma \ref{di_weakclosed}.

\subsubsection{Spectral decomposition}

Let $(Q_n)_n$ be any minimizing sequence for $E_{\mathbf{t}X^{(j_0-1/2)}}^{nr}$ for $j_0\in \tfrac{1}{2}+\mathbb{Z}_+$. 

Thanks to the upper bound, $\text{Dim}\,\mathrm{Ker}(Q_n-1)=1$, as shown in Subsection \ref{di_subscritic}.

There exist a \emph{non-increasing} sequence $(\la_{j;n})_{j\in\mathbb{N}}\in\ell^2$ of eigenvalues and an orthonormal family $\mathbf{B}_n$ of $\mathrm{Ran}\, Q_n$: 
\begin{equation}\label{di_basen}
\mathbf{B}_n:=(\psi_n,\Cha\psi_n)\cup (e_{j;n}^a,e_{j;n}^b,\Cha e_{j;n}^a,\Cha e_{j;n}^b),\ \PP \psi_n=\PP e_{j;n}^{\star}=0,\ \star\in\{a,b\},
\end{equation}
such that the following holds. We omit the index $n$.

\noindent 1. For any $j$, the vector spaces $V_{j;n}^\star:=\PPh(e_{j;n}^\star)$ are irreducible, and so is $V_{0;n}:=\PPh(\psi_n)$. 

\noindent 2. That last one is of type $(\ell_0,\eps(\mathbf{t}))$ (see Notation \ref{di_eps_t}).

\noindent 3. Moreover for any $j\in \mathbb{N}$ we write:
\begin{subequations}\label{di_formtrial}
\begin{equation}\label{di_formtrial1}
 \begin{array}{| l}
   e_{-j}^a:=-\Cha e_{j}^b\text{\ and\ } e_{-j}^b:=\Cha e_j^a,\\
   V_{-j}^a:=\PPh\,e_{-j}^a\text{\ and\ }V_{-j}^b:=\PPh\,e_{-j}^b.
 \end{array}
\end{equation}

\begin{equation}\label{di_formtrial2}
 \begin{array}{rll}
  f_{j}^\star&:=& \sqrt{\tfrac{1-\la_j}{2}} e_{-j}^\star+\sqrt{\tfrac{1+\la_j}{2}}e_{j}^\star,\\
  f_{-j}^\star&:=& -\sqrt{\tfrac{1+\la_j}{2}}e_{-j}^\star+\sqrt{\tfrac{1+\la_j}{2}} e_{j}^\star,
 \end{array}
\end{equation}
and
\begin{equation}\label{di_formtrial22}
 \forall\,j\in \mathbb{Z}^*,\ F_j^\star:=\PPh(f_j^\star).
\end{equation}
The trial state $Q_n$ has the following form.
\begin{equation}\label{di_formtrial3}
\left\{\begin{array}{rll}
 Q_n&=&\text{Proj}\,V_{0,n}-\text{Proj}\,\Cha V_{0,n}+\ssum_{j\ge 1}\la_jq_{j;n} \\
 q_{j;n}&=&\text{Proj}\,F_{j}^a-\text{Proj}\,F_{-j}^a+\text{Proj}\,F_{j}^b-\text{Proj}\,F_{-j}^b.
 \end{array}
 \right.
\end{equation}
\end{subequations}

\begin{remark}\label{di_diag_extrac}
Thanks to the cut-off the sequences $(\psi_n)_n$ and $(e_{j;n})_n$ are $H^1$-bounded. Up to translation and extraction ($(n_k)_k\in\mathbb{N}^{\mathbb{N}}$ and $(x_{n_k})_k\in(\mathbb{R}^3)^{\mathbb{N}}$), we can assume that the weak limit of $(\psi_n)_n$ is non-zero (if it were then there would hold $E_{j_0,\eps(\mathbf{t})}=2m(2j_0+1)$).

We can consider the weak limit of each $(e_n)$: by means of a diagonal extraction, we assume that all the $(e_{j,n_{k}}(\cdot -x_{n_k}))_k$ and $(\psi_{j,n_k}(\cdot-x_{n_k}))_k$, converge along the same subsequence $(n_k)_k$. We also assume that
\begin{equation}\label{di_spec_conv}
 \forall\,j\in\mathbb{N},\ \la_{j,n_k}\to\mu_j,\ (\mu_j)_j\in\ell^2,\ (\mu_j)_j\text{\ non-increasing},
\end{equation}
and that the above convergences also hold in $L^2_{\text{loc}}$ and almost everywhere.
\end{remark}


\subsection{Upper bound and rough lower bound of $E_{j_0,\pm}$}\label{di_subscritic}
We aim to prove the upper bound of Proposition \ref{di_est}. The method will also give a rough lower bound of $E_{j_0,\pm}$.
\begin{notation}
 We write:
 \[
  C(j_0):=j_0^2\underset{-j_0\le m\le j_0}{\sup}\nlp{\infty}{\Psi_{m,j_0\pm \tfrac{1}{2}}}^4,
 \]
where the functions $\Psi_{m,j_0\pm\tfrac{1}{2}}$ are defined in \cite[p. 125]{Th}: they are the upper or lower spinors of the $\Phi^{\pm}_{m,\kappa_{j_0}}$'s.
\end{notation}

For $E_{j_0,\eps(\mathbf{t}})$, we only consider $\mathbf{t}\in\{(1,0);(0,1)\}$ and $\eps(\mathbf{t})$ is defined in Notation \ref{di_eps_t}.


\noindent -- We consider trial state of the following form:
\[
 Q=\text{Proj}\,\PPh(\psi)-\text{Proj}\,\PPh(\Cha\psi),
\]
where $\PPh(\psi)$ is of type $(\ell_0+\tfrac{1}{2},\eps(\mathbf{t}))$ and $\PP\psi=0$. For short, we write
\[
 N_\psi:=\text{Proj}\,\PPh(\psi)\text{\ and\ }N_{\Cha \psi}:=\text{Proj}\,\PPh(\Cha\psi).
\]

The set of these states is written $\mathscr{W}_{\mathbf{t}X^{\ell_0}}^0$. We will prove that the energy of a particular $Q$ gives the upper bound. The BDF energy of $Q\in \mathscr{W}_{\mathbf{t}X^{\ell_0}}^0$ is:
\begin{equation}\label{di_form_no_pol}
 2\ttr\big(|\D|N_{\psi}\big)-\alpha\nqq{N_\psi}^2-\alpha \mathfrak{Re}\,\ttr\big(N_\psi R[N_{\Cha \psi}]\big).
\end{equation}

\noindent -- We will study the non-relativistic limit $\alpha\to 0$.

\noindent -- To get an upper bound, we choose a specific trial state in $\mathscr{W}_{\mathbf{t}X^{\ell_0}}$, the idea is the same as in \cite{sok,pos_sok}: the trial state is written in \eqref{di_trial_non_rel_end}. Before that, we precise the structure of elements in $\mathscr{W}_{\mathbf{t}X^{\ell_0}}^0$.

\paragraph{Minimizer for $E_{\mathbf{t}X^{\ell_0}}^{nr}$} By an easy scaling argument, there exists a minimizer for the non-relativistic energy $E_{\mathbf{t}X^{\ell_0}}^{nr}$ \eqref{di_non_rel_w}. The scaling argument enables us to say that this energy is negative. Then it is clear that a minimizing sequence converges to a minimizer $\ov{\G}$, up to extraction.
Writing
\[
 H_{\ov{\G}}:=-\Delta-R_{\ov{\G}},
\]
this minimizer satisfies the self-consistent equation
\[
 \big[H_{\ov{\G}}, \ov{\G}\big]=0.
\]
This comes from Remark \ref{di_prec_non_rel}. In particular, $H_{\ov{\G}}$ restricted to $\mathrm{Ran}\,\G$ is a homothety by some $-e^2<0$, so
\[
 \forall\,\psi\in\mathrm{Ran}\,\ov{\G},\ \nlp{2}{\psi}=1,\ \nlp{2}{\Delta \psi}\le \nlp{2}{R_{\ov{\G}}\psi}\apprle \nqq{\ov{\G}}\nlp{2}{\,|\nabla|^{1/2}\psi},
\]
and we get
\[
 \nlp{2}{\Delta \psi}^{3/4}\apprle \nqq{\ov{\G}}\ i.e.\ \nlp{2}{\Delta \psi}\apprle \nqq{\ov{\G}}^{4/3}\apprle (2j_0+1)^{2/3}.
\]
The last estimate comes from a simple study of a minimizer for $E_{\mathbf{t}X^{\ell_0}}^{nr}$: we have
\[
 \ttr\big(-\Delta \ov{\G}\big)-\frac{\pi}{2}\ttr\big(|\nabla|\ov{\G}\big)\le \mathcal{E}_{nr}(\ov{\G})<0,
\]
thus $\ttr\big(-\Delta \ov{\G}\big)\apprle j_0^2$ and $ \ttr\big((-\Delta)^2\ov{\G}\big)\apprle j_0^{5/2}.$

We end this bootstrap argument at $\nlp{2}{|\nabla|^{3}\psi}$ for $\psi\in\mathrm{Ran}\,\psi$: we have
\begin{align*}
 |\nabla|^3\psi&=\frac{-\Delta}{e^2-\Delta}\Big([|\nabla|,R_{\ov{\G}}]\psi+R_{\ov{\G}}\psi\Big),\\
 \nlp{2}{\,|\nabla|^3\psi}&\apprle \ns{2}{\Delta \ov{\G}}+\ns{2}{\nabla \ov{\G}}\apprle j_0^{5/2}.
\end{align*}

\medskip

\paragraph{Trial state} We take the following trial state. First, let $\ov{\G}=\text{Proj}\,ra_0(r)\Psi_{j_0,j_0+\eps(\mathbf{t})\tfrac{1}{2}}$ be a minimizer for $E_{\mathbf{t}X^{\ell_0}}^{nr}$. We form
\begin{equation}\label{di_trial_non_rel_1}
\ov{N}_+:= \text{Proj}\,\PPh\, \PPP U_{\la^{-1}}(ra_0(r)\Phi^{+}_{j_0,\eps(\mathbf{t})(j_0+\tfrac{1}{2})})
\end{equation}
where we recall that
\[
 \la:=\frac{g'_1(0)^2}{\alpha m}\text{\ and\ }U_a \phi(x):=a^{3/2} \phi(ax),\ a>0.
\]
This corresponds to dilating $\ov{\G}$ by $\la^{-1}$ and projecting the range of the dilation onto $\mathrm{Ran}\,\PPP$. Of course $ \G\in \mathfrak{S}_1(L^2(\RR,\mathbb{C}^2))$ is embedded in $\mathfrak{S}_1(L^2(\RR,\mathbb{C}^2\times\mathbb{C}^2))$ as follows:
\[
 \ov{\G}\mapsto \begin{pmatrix} \ov{\G} & 0 \\ 0 & 0\end{pmatrix}\in \mathfrak{S}_1(L^2(\RR,\mathbb{C}^2\times\mathbb{C}^2)).
\]

Then we define
\begin{equation}\label{di_trial_non_rel_2}
\ov{N}_-:=\Cha \ov{N}_-\Cha^{-1}=\Cha \ov{N}_-\Cha.
\end{equation}
Our trial state is
\begin{equation}\label{di_trial_non_rel_end}
 \ov{N}:=\ov{N}_+-\ov{N}_-.
\end{equation}

\paragraph{Upper bound for $E_{j_0,\pm}$} We compute $\mathcal{E}^0_{\text{BDF}}(\ov{N})$.

Before that, we study a general projector $\text{Proj}\,\PPh\,\psi$ where $\PP\psi=0$ and $\PPh\,\psi$ irreducible of type $(j_0,\eps(\mathbf{t}))$.

As an element of $\mathrm{Ran}\,\PPP$, the wave function $\psi$ can be written
\[
 \psi=\PPP \begin{pmatrix}\ph\\ 0\end{pmatrix}.
\]
As it spans an irreducible representation of type $(j_0,\eps(\mathbf{t}))$, we can choose
\[
 \forall\, x=r\om_x\in\RR,\ \ph(x):= ia(r)\Psi_{j_0+\eps(\mathbf{t})\tfrac{1}{2}}^{j_0}(\om_x),\ a(r)\in L^2\big((0,\infty),r^2dr\big),
\]
where we used notations of \cite[p. 126]{Th}. This corresponds to taking
\[
 \psi:=\PPP ra(r)\Phi^+_{j_0,\eps(j_0+\tfrac{1}{2})},\ \eps=\eps(\mathbf{t}).
\]

We recall the following formulae of \cite[pp. 125-127]{Th} (with $\boldsymbol{\om}:x\mapsto\tfrac{x}{|x|}$)
\begin{equation}\label{di_form_th_125}
\begin{array}{l}
 -i\boldsymbol{\alpha}\cdot \nabla=-i(\boldsymbol{\alpha}\cdot \boldsymbol{\om})\partial_r+\frac{i}{r}(\boldsymbol{\alpha}\cdot \boldsymbol{\om})(2\Spbf\cdot \Lbf),\\
 \big\{\Spbf\cdot \Lbf,\boldsymbol{\alpha}\cdot \boldsymbol{\om}\big\}=-\boldsymbol{\alpha}\cdot \boldsymbol{\om}\text{\ and\ }i\boldsymbol{\sigma}\cdot \boldsymbol{\om}\Psi^{m_j}_{j\pm \tfrac{1}{2}}=\Psi^{m_j}_{j\mp\tfrac{1}{2}}.
 \end{array}
\end{equation}
This gives
\begin{equation}\label{di_form_en_trial1}
\begin{array}{rcl}
  \PPP a(r)\Phi^+_{m,\eps(\mathbf{t})(j_0+\tfrac{1}{2})}&=&\dfrac{1}{2}\begin{pmatrix}i\big(1+\frac{g_0(|\nabla|)}{|\D|} \big)a(r) \Psi^m_{j_0+\eps\tfrac{1}{2}}\\ \frac{g_1(|\nabla|)}{|\D||\nabla|}\big(\partial_r (a(r))+\eps(j_0+\tfrac{1}{2})\tfrac{a(r)}{r}\big)\Psi^m_{j_0-\eps\tfrac{1}{2}}\end{pmatrix},\\
  										  &=:&\begin{pmatrix} ia_{\uparrow}(r)\Psi^m_{j_0+\eps\tfrac{1}{2}}\\ a_{\downarrow}(\eps,j_0;r)\Psi^m_{j_0-\eps\tfrac{1}{2}}\end{pmatrix}.
\end{array}
\end{equation}
We write $ \mathrm{Op}:=\frac{g_1(|\nabla|)}{|\D||\nabla|}:$ the following holds.
\begin{equation}\label{di_form_en_trial2}
\begin{array}{rcl}
 \Big| \ttr\big(N_\psi R[N_{\Cha \psi}]\big)\Big|&\apprle &j_0^2\sup_{m}\nlp{\infty}{\Psi^m_{j_0\pm\tfrac{1}{2}}}^2\ncc{\,|a_{\uparrow}a_{\downarrow}(\eps,j_0,\cdot)|}^2\\
    &\apprle&  C(j_0)D\Big(|a_{\uparrow}|^2; |\mathrm{Op}\cdot\partial_r (a(r))|^2+j_0^2|\mathrm{Op}\cdot r^{-1}a(r)|^2\Big),\\
    &\apprle& C(j_0) \psh{|\nabla|\psi}{\psi}\nlp{2}{\nabla\psi}^2=: \mathcal{R}em_0(j_0,\psi).
\end{array}
 \end{equation}

In fact, we have $\ttr\big(N_\psi R[N_{\Cha \psi}] \big)\ge 0$ by direct computation.

Let us deal with $\nqq{N_\psi}^2$.
\begin{notation}
 We write $\Pup$ the projection onto the upper part of $\mathbb{C}^2\times\mathbb{C}^2$ and $\Pdow$ the projection onto the lower part. That is: $\Pup \psi$ has no lower spinor and the same upper spinor as $\psi$.
 
\end{notation}

Similarly, 
\begin{align*}
 \nqq{N_\psi}^2-\nqq{\Pup N_\psi \Pup}^2&=\ttr\big(\Pup N_\psi \Pdow R_{N_\psi}\big)+\ttr\big(\Pdow N_\psi \Pup R_{N_\psi}\big)\\
               &\ \ +\nqq{\Pdow N_\psi \Pdow},\\
               &\apprle \mathcal{R}em(j_0,\psi)+C(j_0)\nlp{2}{\nabla \psi}^2\nlp{2}{\tfrac{|\nabla|^{3/2}}{|D_0|} \psi}^2,\\
               &=:\mathcal{R}em_1(j_0,\psi).
\end{align*}
For the trial state \eqref{di_trial_non_rel_end}, this gives:
\begin{align*}
 \nqq{\ov{N}_+}^2&=\nqq{\Pup N_\psi \Pup}^2+\mathcal{O}\Big( C(j_0)\big( \alpha^3j_0+\alpha^5j_0^{5/3}\big)\Big)\\
		 &=\frac{\alpha m}{g'_1(0)^2}\nqq{\ov{\G}}^2(1+\mathcal{O}(\nlp{2}{\nabla \psi}^2))\\
		 &\ \ \ +\mathcal{O}\big( \nlp{2}{\tfrac{\Delta}{1-\Delta} \psi}^2(\nlp{2}{\tfrac{|\nabla|^{5/2}}{1-\Delta} \psi}+\nlp{2}{\nabla \psi}^2)\big),\\
		 &=\frac{\alpha m}{g'_1(0)^2}\nqq{\ov{\G}}^2\\
		 &+\mathcal{O}\Big[C(j_0)\Big(\alpha^3j_0^{5/3}+ \underset{0\le s\le 1}{\inf}(\alpha^{4s}j_0^{4s/3})\big(\alpha^2j_0^{2/3}+\underset{2^{-1}\le s\le 1}{\inf}(\alpha^{4s}j_0^{4s/3})\big) \Big)\Big].
\end{align*}

We compute the kinetic energy as in \cite{sok,pos_sok}: we get
\begin{align*}
 \ttr\big(|\D| \ov{N}_+\big)&=\frac{\alpha^2m}{g'_1(0)^2}\ttr\big(-\Delta \ov{\G}\big)\big(1+K\alpha\big)+\mathcal{O}\big(\alpha^4\ttr\big((\Delta)^2\ov{\G}\big)\big),\\
                            &=\frac{\alpha^2m}{g'_1(0)^2}\ttr\big(-\Delta \ov{\G}\big)+\mathcal{O}\big(\alpha^3 j_0+\alpha^4j_0^{5/2}\big).
\end{align*}
This proves
\begin{equation}
\begin{array}{|l}
E_{j_0,\eps(\mathbf{t})}\le 2m(2j_0+1)+\frac{\alpha^2m}{g'_1(0)^2}E_{\mathbf{t}X^{\ell_0}}^{nr}+\mathcal{O}\big(\varrho(\alpha,j_0)\big)\\
\varrho(\alpha,j_0):=\alpha^3 j_0+\alpha^4j_0^{5/2}+C(j_0)\Big(\alpha^3j_0^{5/3}+ \underset{0\le s\le 1}{\inf}(\alpha^{4s}j_0^{4s/3})\big(\alpha^2j_0^{2/3}+\underset{2^{-1}\le s\le 1}{\inf}(\alpha^{4s}j_0^{4s/3})\big) \Big).
\end{array}
\end{equation}


First, by Kato's inequality \eqref{di_kato}, we have
\[
 \nqq{N_\psi-N_{\Cha \psi}}^2\le \frac{\pi}{2}\ttr\big(|\nabla| (N_\psi+N_{\Cha \psi})\big)=\pi\ttr\big(|\nabla|N_{\psi}\big).
\]
So
\[
 \mathcal{E}^0_{\text{BDF}}(Q)\ge 2\Big(\ttr\big(|\D|N_{\psi}\big)-\alpha\frac{\pi}{2}\ttr\big(|\nabla|N_{\psi}\big)\Big)=:2\big((2j_0+1)m+\mathcal{F}(N_\psi)\big).
\]
As $\alpha$ tends to $0$, a minimizer over $\mathscr{W}_{\mathbf{t}X^{\ell_0}}^0$ should be localized in Fourier space around $0$. Indeed, for $\alpha,L$ sufficiently small, we have
\[
 \forall\,p\in B(0,\La),\ \ed{p}-m=\frac{g_0(p)^2-m^2+g_1(p)^2}{\ed{p}+m}\ge \frac{p^2}{2\nlp{\infty}{g_0}|D_0|},
\]
and for any $0<s\le 2$:
\[
 \frac{p^2}{2\nlp{\infty}{g_0}|D_0|}\ge s\frac{\alpha\pi}{2}|p|\iff |p|\ge \frac{\alpha s\pi \nlp{\infty}{g_0}}{\sqrt{1-(\alpha s\pi\nlp{\infty}{g_0})^2}}=:\vartheta_{s}.
\]
We get
\[
2\mathcal{F}\big( \Pi_{\vartheta_1} N_{\psi} \Pi_{\vartheta_1}\big)\le \mathcal{E}^0_{\text{BDF}}(Q)-2(2j_0+1)m.
\]
By Cauchy-Schwartz inequality, we get a rough lower bound
\[
 \ttr\big(-\Delta \Pi_{\vartheta_1} N_{\psi} \Pi_{\vartheta_1} \big)\apprle \alpha^2(2j_0+1)\text{\ and\ }\mathcal{E}^0_{\text{BDF}}(Q)-2(2j_0+1)m\apprge -\alpha^2(2j_0+1).
\]
For an almost minimizer $Q$, the same argument shows that
\begin{equation}\label{di_alm_min}
\ttr\big(\frac{-\Delta}{|\D|}Q^2\big)\apprle \alpha^2 (2j_0+1).
\end{equation}

A precise lower bound is obtained once we know that there exists a minimizer $\ov{P}_{j_0}$. This state satisfies the self-consistent equation \eqref{di_eq_min}: see Subsection \ref{di_low_bound}.





\begin{remark}
 The same method can be used to get an upper bound of $E_{p(X)}^{nr}$ for any $p(X)=\sum_{\ell=0}^{\ell_0}\mathbf{t}_{\ell}X^\ell$. By scaling we have $E_{p(X)}^{nr}<0.$
\end{remark}

\subsection{Strategy of the proof: the para-positronium}

The method is more subtle because $\mathscr{M}_{\mathscr{I}}$ has only one connected component. We first consider the subset $\mathscr{M}_{\mathscr{I}}^{1}$ defined by:
\begin{equation}\label{di_trial_isym}
 \mathscr{M}_{\mathscr{I}}^{1}=\big\{P_\psi:=\PP+\ket{\psi}\bra{\psi}-\ket{\Isym\psi}\bra{\Isym\psi},\ \psi\in\mathbb{S}\mathrm{Ran}\,\PPP\big\}.
\end{equation}

\begin{lemma}\label{di_infimum_1}
 Let $F_{\mathscr{I}}$ be the infimum of the BDF energy over $\mathscr{M}_{\mathscr{I}}^{1}$. Then we have
 \begin{equation}
  F_{\mathscr{I}}\ge 2m-\alpha^2\frac{E_{\mathrm{PT}}(1) m}{g'_1(0)^2}+\mathcal{O}(\alpha^3).
 \end{equation}
\end{lemma}

We will prove the existence of a critical point in the neighbourhood of $\mathscr{M}_{\mathscr{I}}^{1}$ \emph{via} a mountain pass argument. In this part, we aim to prove the following Proposition.
\begin{proposition}\label{di_para_method}
\noindent 1. In the regime of Theorem \ref{di_main}, there exists a bounded sequence in $\mathscr{M}_{\mathscr{I}}-\PP$ of almost critical points: $(Q_n=P_n-\PP)_n$ such that
\[
\underset{n\to+\infty}{\lim}\ns{2}{\nabla \mathcal{E}^0_{\text{BDF}}(P_n)}=0\mathrm{\ with\ }\mathcal{E}^0_{\text{BDF}}(Q_n)= 2m-\frac{\alpha^2 m}{g'_1(0)^2}E_{\text{PT}}(1)+\mathcal{O}(\alpha^3).
\]
Furthermore, for sufficiently big $n$, there exists $\psi_{a;n}$ such that
\[
\mathbb{C}\psi_{a;n}=\mathrm{Ran}\,P_n\cap \mathrm{Ran}\,\chi_{(0,+\infty)}\big(D_{Q_n}^{(\La)}-\nabla \mathcal{E}^0_{\text{BDF}}(P_n)\big)
\]
and $P_n=\chi_{(-\infty,0)}\big(D_{Q_n}^{(\La)}-\nabla \mathcal{E}^0_{\text{BDF}}(P_n)\big)+\ket{\psi_{a;n}}\bra{\psi_{a;n}}-\ket{\Isym\psi_{a;n}}\bra{\Isym \psi_{a;n}}.$

\noindent 2. Up to a subsequence and up to translation the sequence tends to a critical point $Q_{\infty}$ of $\mathcal{E}^0_{\text{BDF}}$ in $\mathscr{M}_{\mathscr{I}}-\PP$.

Moreover, writing $\ov{P}=Q_\infty+\PP$, there exists $0<\mu<m$ and $\psi_a\in \mathbb{S}\,\hl$ such that
\begin{equation}
 \left\{\begin{array}{ccl}
          \ov{P}&=&\chi_{(-\infty,0)}(D_{Q_\infty}^{(\La)})+\ket{\psi_a}\bra{\psi_a}-\ket{\Isym \psi_a}\bra{\Isym \psi_a},\\
          \mathbb{C}\psi_a&=&\mathrm{Ker}\big(D_{Q_\infty}^{(\La)}-\mu\big),\\
          \inf\sigma(|D_{Q_\infty}^{(\La)}|)&=&\mu.
        \end{array}
\right.
\end{equation}

\end{proposition}

\paragraph{Proof of Proposition \ref{di_para_method}: first part}

For any $\psi\in \mathbb{S}\mathrm{Ran}\,\PPP$, we define:
\begin{equation}\label{di_def_c(t)}
 c_{\psi}:\begin{array}{rcl}
      [0,1] &\longrightarrow&  \mathscr{M}_{\mathscr{I}}-\PP\\
      s&\mapsto& \ket{\sin(\pi s)\psi+\cos(\pi s)\Isym \psi}\bra{\sin(\pi s)\psi+\cos(\pi s)\Isym \psi}-\ket{\Isym \psi}\bra{\Isym \psi}.
   \end{array}
\end{equation}

\begin{remark}\label{di_remark_cross}
The loop $c_\psi+\PP$ crosses $\mathscr{M}_{\mathscr{I}}^{1}$ at $t_0=\tfrac{1}{2}$ where the BDF energy is maximal:
\[
 \underset{s\in[0,1]}{\sup}\mathcal{E}^0_{\text{BDF}}(c(s)).
\]
Indeed, there holds
\[
 \mathcal{E}^0_{\text{BDF}}(c(s))=2\sin(\pi s)^2\psh{|\D|\psi}{\psi}-\alpha\sin(\pi s)^2\big[D\big(|\psi|^2,|\psi|^2\big)+\cos(2\pi s)D\big(\psi^*\Isym \psi, \psi^*\Isym\psi\big)\big], 
\]
and the derivative with respect to $s$ is:
\[
\begin{array}{l}
 \frac{d}{d s} \mathcal{E}^0_{\text{BDF}}(c(s_0))=2\pi\sin(2\pi s_0)\Big(\psh{|\D|\psi}{\psi}-\frac{\alpha}{2}\big[D\big(|\psi|^2,|\psi|^2\big)\\
 \ \ \ \ \  \ \ \ \ \  \ \ \ \ \  \ \ \ \ \ +(\sin(\pi s_0)^2-\tfrac{1}{2}\cos(2\pi s_0)) \alpha D\big(\psi^*\Isym \psi, \psi^*\Isym\psi\big)\big]\Big).
 \end{array}
\]
For sufficiently small $\alpha$, this quantity vanishes only at $2\pi s_0\equiv 0[\pi]$.
\end{remark}

What happens when we apply the gradient flow $\Phi_{\text{BDF},t}$ of the BDF energy ? The loop $c_{\psi}$ is transformed into $c_{t}:=\Phi_{\text{BDF},t}(c_\psi)$ and we still have 
\[
 c_t(s=0)=c_t(s=1)=0.
\]
This follows from the fact that $\PP$ is the global minimizer of $\mathcal{E}^0_{\text{BDF}}$.

We recall that for all $s\in[0,1]$, the function $c_t(s)$ satisfies the equation
\[
 \forall\,t_0\in\mathbb{R}_+,\ \frac{d}{dt}(c_{t_0}(s))=-\nabla \mathcal{E}^0_{\text{BDF}}(c_{t_0}(s))\in \text{T}_{c_{t_0}(s)+\PP}\mathscr{M}_{\mathscr{I}}.
\]

The non-trivial result holds.
\begin{lemma}\label{di_non_triv}
 Let $P_\psi\in\mathscr{M}_{\mathscr{I}}^1$ be a state whose energy is close to the infimum $F_{\mathscr{I}}$:
 \[
  \mathcal{E}^0_{\text{BDF}}\big(P_\psi\big)<F_{\mathscr{I}}+\alpha^3.
 \]
Let $c_\psi$ be the loop associated to $\psi$ (see \eqref{di_def_c(t)}) and $c_t:=\Phi_{\text{BDF},t}(c_\psi)$. Then for all $t\in\mathbb{R}_+$, the loop $c_t$ crosses the set $\mathscr{M}_{\mathscr{I}}^{1}$ at some $\wt{s}(t)\in(0,1)$. 
\end{lemma}

\begin{lemma}\label{di_ex_critic}
 Let $(c_t)_{t\ge 0}$ be the family of loops defined in Lemma \ref{di_non_triv} and let $(s(t))_{t\ge 0}$ be a family of reals in $(0,1)$ such that
 \[
  \forall\,t\ge 0,\ \mathcal{E}^0_{\text{BDF}}\big(c_t(s(t))\big)=\underset{s\in[0,1]}{\sup}\mathcal{E}^0_{\text{BDF}}(c_t(s)).
 \]
Then there exists an increasing sequence $(t_n)_{n\in\mathbb{N}}$ the sequence $(c_{t_n}(s(t_n)))_{n\ge 0}$ satisfies the first point of Proposition \ref{di_para_method}
\end{lemma}

We prove Lemmas \ref{di_infimum_1} and \ref{di_non_triv} in Subsection \ref{di_fait_ch}. We assume they are true to prove Lemma \ref{di_ex_critic} and Proposition \ref{di_para_method}.

\begin{remark}
The proof of Lemma \ref{di_non_triv} uses an index argument. We kept it elementary but it is possible to rephrase it in terms of the Maslov index \cite{LagGrass} once we notice that $\Isym$ induces a symplectic structure and that the projectors in $\mathscr{M}_{\mathscr{I}}$ are Lagrangians (see Remark \ref{Lagrangians}).
\end{remark}

\subparagraph{Spectral decomposition of $P_n$}

We define
\[
 F_1:=\liminf_{t\to +\infty}\mathcal{E}^0_{\text{BDF}}(c_t(s(t)))=\liminf_{t\to+\infty}\underset{s\in[0,1]}{\sup}\mathcal{E}^0_{\text{BDF}}(c_t(s)).
\]
We assume $(t_n)_{n\ge 0}$ is a minimizing sequence for $F_1$.

We may assume that $\lim_{n\to+\infty}t_n=+\infty$. 

\noindent -- First we prove that along the path $c_t$ the gradient $\nabla \mathcal{E}^0_{\text{BDF}}$ (see \eqref{di_defgradient}) is bounded in $\mathfrak{S}_2$. Indeed, for all $P=Q+\PP\in\mathscr{M}$, we write
\[
 \wt{Q}:=P-\chi_{(-\infty,0)}\big(\Pi_\La D_Q\Pi_\La\big),
\]
We recall that $D_Q^{(\La)}:=\Pi_\La D_Q \Pi_\La$:
\begin{equation}\label{di_form_grad}
\begin{array}{rcl}
 \nabla \mathcal{E}^0_{\text{BDF}}(P)&=&\big[\big[D_Q^{\La},P\big],P\big]=\big\{|D_Q^{(\La)}|;\wt{Q}\big\}-2\wt{Q} D_{Q}^{(\La)} \wt{Q},\\
 \lVert \nabla \mathcal{E}^0_{\text{BDF}}(P)\rVert_{\mathfrak{S}_2}&\apprle&\ns{2}{\wt{Q}}\ed{\La}\Big[(1+\ns{2}{Q})(1+\ns{2}{\wt{Q}})\Big]\\
                             &\apprle& K(\La,F_1+\alpha^3).
 \end{array}
\end{equation}
We have used the Cauchy expansion \eqref{di_cauchy20} to get an expression
\[
 \chi_{(-\infty,0)}\big(D_Q^{(\La)}\big)-\PP=\ssum_{k=1}^{+\infty}\alpha^k M_k[Y[Q]]
\]
where $M_k[Y[Q]]$ is a polynomial function of $\pi_\La R_{Q}\Pi_\La$ of degree $k$. We refer the reader to these papers or to \eqref{di_cauchy0}-\eqref{di_cauchy_est0} above for more details.

From formula \eqref{di_form_grad} and Remark \ref{di_diff_ch} we see that the gradient, as a function of $Q$ is \emph{locally Lipschitz}, at least in some ball $\{Q_0:\,\ns{2}{|\D|^{1/2} Q_0}\le C_0\}$ in which there holds
\[
 \inf \sigma\big(|D_{Q_0}^{(\La)}|\big)\ge K(C_0),
\]
where $C_0$ is some constant. The Lipschitz constant depends on the constant $C_0$ and in the present case, we can take $C_0\apprle 1$.

Let us prove that
\begin{equation}\label{di_grad_zero}
 \lim_{n\to+\infty}\ns{2}{\nabla \mathcal{E}^0_{\text{BDF}}(c_{t_n}(s(t_n))) }=0.
\end{equation}
If not, the $\limsup$ is bigger than some $\eta>0$ and then we get a contradiction when we consider $n_0$ large enough such that
\[
 |F_1-\mathcal{E}^0_{\text{BDF}}(c_{t_{n_0}} (s(t_{n_0})))|\ll \eta\text{\ and\ }\ns{2}{\nabla \mathcal{E}^0_{\text{BDF}}(c_{t_{n_0}}(s(t_{n_0})))}\ge \frac{\eta}{2},
\]
because
\[
 \forall\,\tau>0,\ \mathcal{E}^0_{\text{BDF}}(c_{t_{n_0}+\tau}(s(t_{n_0})))-\mathcal{E}^0_{\text{BDF}}(c_{t_{n_0}}(s(t_{n_0})))=-\dint_0^{\tau}\ns{2}{\nabla \mathcal{E}^0_{\text{BDF}}(c_{t_{n_0}+u})(s_{t_{n_0}}) }^2du.
\]

\noindent -- We recall that the gradient at $P\in\mathscr{M}$ is the "projection" of the mean-field operator onto the tangent plane $\text{T}_{P}\mathscr{M}$, in the sens that
\[
\begin{array}{l}
 \forall\,v\in \text{T}_{P}\mathscr{M},P D_Q (1-P)\in\mathfrak{S}_1\text{\ and\ }\\
 \ \ \ \ \ \ \ \ \ \ \ \ \ttr\big(P D_Q (1-P) v+(1-P)D_Q P v\big)=\ttr\big( \nabla \mathcal{E}^0_{\text{BDF}} \big)
 \end{array}
\]

\begin{notation}
For short, we write
\[
 Q_n:=c_{t_n}\big(s(t_n)\big)\text{\ and\ }P_n:=Q_n\text{\ and\ }v_n:=\nabla \mathcal{E}^0_{\text{BDF}}(Q_n).
\]
Moreover, we write
\[
\wt{D}_{Q_n}:=D_{Q_n}-v_n\text{\ and\ } \wt{\boldsymbol{\pi}}_{-;n}:=\chi_{(-\infty,0)}\big(D_{Q_n}^{(\La)}-v_n\big).
\]

\end{notation}

We have shown that $\lim_{n\to+\infty}\ns{2}{v_n}=0.$

But as $v_n$ is an element of the tangent plane $\text{T}_{P_n}\mathscr{M}$, we have
\[
 \big[ \big[v_n,P_n\big] ,P_n\big]=P_n v_n(1-P_n)+(1-P_n)v_nP_n=v_n
\]
thus
\[
 \big[ \big[ D_{Q_n}^{(\La)}-v_n,P_n\big] ,P_n\big]=0.
\]
Equivalently, we have
\begin{equation}\label{di_comm_alm}
 \big[\wt{D}_{Q_n}^{(\La)},P_n \big]=(1-P_n)\wt{D}_{Q_n}^{(\La)}P_n-P_n\wt{D}_{Q_n}^{(\La)}(1-P_n)=0.
\end{equation}

Thus the projector $P_n$ commutes with the distorted mean-field operator $\wt{D}_{Q_n}$. We recall that
\[
\lim_n \ns{2}{\wt{D}_{Q_n}^{(\La)}-D_{Q_n}^{(\La)}}=0,
\]
and thus up to taking $n$ big enough, we can neglect the distortion $v_n$: all its Sobolev norms tend to zero as $n$ tends to infinity \emph{thanks to the cut-off}.

\noindent -- Thanks to Lemma \ref{di_infimum_1} we have the following energy condition:
\[
 2m+\mathcal{O}(\alpha^2)\le F_1\le \mathcal{E}^0_{\text{BDF}}(Q_n)\le F_1+\alpha^3=2m+\mathcal{O}(\alpha^2).
\]
Using the Cauchy expansion \eqref{di_cauchy0}-\eqref{di_cauchy_est0}, we have
\[
 \ns{2}{\,|\D|^{1/2}(\wt{\boldsymbol{\pi}}_{-;n}-\PP)}\apprle \sqrt{L\alpha}\nqq{Q_n}\apprle \sqrt{L\alpha}.
\]

Thus we get
\[
 \big|\ns{2}{Q_n}-\ns{2}{P_n-\wt{\boldsymbol{\pi}}_{-;n}}\big|\le \ns{2}{\PP-\wt{\boldsymbol{\pi}}_{-;n}}\apprle \sqrt{L\alpha}.
\]
As $\wt{D}_{Q_n}$ and $P_n$ commutes, then necessarily $\ns{2}{P_n-\wt{\boldsymbol{\pi}}_{-;n}}^2$ is an integer equal to twice the dimension of $\mathrm{Ran}\,P_n\cap \mathrm{Ran}\,(1-\wt{\boldsymbol{\pi}}_{-;n})$.

But we know that
\[
  m\ns{2}{Q_n}^2\le \ttr\big(|\D|Q_n^2\big)\le \frac{1}{1-\alpha \tfrac{\pi}{4}}\mathcal{E}^0_{\text{BDF}}(Q_n)\le \frac{2m}{1-\alpha\frac{\pi}{4}}=2m+\mathcal{O}(\alpha).
\]
Then the above dimension is lesser than $1$ and it cannot be $0$ because of the energy condition 
\[
 \mathcal{E}^0_{\text{BDF}}(Q_n)\ge F_{\mathscr{I}}\ge 2m-K\alpha^2\gg \sqrt{L\alpha}.
\]
This proves the first part of Proposition \ref{di_para_method}. We have $\mathrm{Ran}\,P_n\cap \mathrm{Ran}\,(1-\wt{\boldsymbol{\pi}}_{-;n})=\mathbb{C}\psi_{a;n}$ where $\psi_{a;n}$ is unitary. It is an eigenvector for $\wt{D}_{Q_n}^{(\La)}$ with eigenvalue $\mu_n$. From the equality:
\[
\mathcal{E}^0_{\text{BDF}}(Q_n)=\mathcal{E}^0_{\text{BDF}}(\wt{\boldsymbol{\pi}}_{-;n}-\PP)+2\mu_n-\frac{\alpha}{2}\diint\frac{|\psi_{a;n}\wedge \Isym \psi_{a;n}(x,y)|^2}{|x-y|}dxdy,
\]
we get $0<\mu_n<m$. We end the proof as follows.

\paragraph{Proof of Proposition \ref{di_para_method}: second part}
We follow the method of \cite{pos_sok}. We recall the main steps and refer the reader to this paper for further details.

\noindent -- The idea is simple: we must ensure that there exists a non-vanishing weak-limit and that this weak-limit is in fact a critical point.

Let us say that $\psi_{a;n}$ is associated to the eigenvalue $\mu_n$. 

\noindent -- The condition of the energy ensures that the sequence $(\psi_{a;n})_n$ does not vanish in the sense that we \emph{do not} have the following:
\[
\forall\,A>0,\ \limsup_n \sup_{x\in\RR}\dint_{B(x,A)}|\psi_{a;n}|^2=0.
\]
Up to translation and extraction of a subsequence, we may suppose that $(Q_n)$ (resp. $(\psi_{a;n})$) converges in the weak topology of $H^1$ to $Q_\infty\neq 0$ (resp. $\psi_{a}\neq 0$). In particular these sequences also converge in $L^2_{loc}$ and \emph{a.e.} We recall that thanks to the cut-off and Kato's inequality \eqref{di_kato} , we have $Q_n\in H^1(\RR\times \RR)$ with
\[
\ns{2}{|D_0|Q_n}^2\le \ed{\La}\ns{2}{|\D|^{1/2}Q_n}^2\le \frac{\ed{\La}}{1-\alpha \pi/4}\sup_n \mathcal{E}^0_{\text{BDF}}(Q_n).
\]
A similar estimate hold for $(\psi_{a;n})$. We also suppose that $\lim_n\mu_n=\mu_{\infty}$.

\noindent -- As shown in \cite{pos_sok}, the operator $R_{Q_n}$ converges in the strong operator topology to $R_{Q_\infty}$. Thanks to the Cauchy expansion \eqref{di_cauchy20}, we also have
\[
\text{s}.\,\lim_n\Big[\chi_{(-\infty,0)}\big(D_{Q_n}^{(\La)}-\nabla \mathcal{E}^0_{\text{BDF}}(P_n)\big)-\PP\Big]=\chi_{(-\infty,0)}\big(D_{Q_\infty}^{(\La)}\big)-\PP.
\]
By that strong convergence, we also have the weak-convergence of $\wt{D}_{Q_n}^{(\La)}\psi_{a;n}$ to $D_{Q_\infty}^{(\La)}\psi_a$ in $L^2$ and it follows that:
\[
D_{Q_\infty}^{(\La)}\psi_a=\mu_{\infty}\psi_a\neq 0.
\]

\noindent -- The condition of the energy ensures that for $\alpha$ sufficiently small, the $\psi_{a;n}$'s are close to a scaled Pekar minimizer: for any $n$, there exists a Pekar minimizer $\wt{\phi}_n$ such that
\[
\lVert\psi_{a;n}-\la^{-3/2}\wt{\phi}_n(\la^{-1}(\cdot))\rVert_{H^1}^2\le \alpha K\text{\ where\ }\la:=\frac{g'_1(0)^2}{\alpha m}.
\]
The constant $K$ depends on the energy estimate of Proposition \ref{di_para_method}.

\noindent -- Thanks to that, for all $n$, $\mu_n$ is an isolated eigenvalue of $\wt{D}_{Q_n}^{(\La)}$, uniformly in $n$: we have 
\[
\mathbb{C}\psi_{a;n}=\mathrm{Ker}\big(\wt{D}_{Q_n}^{(\La)}-\mu_n\big),
\]
and 
\[
\text{dist}\Big(\mu_n;\sigma\big(\wt{D}_{Q_n}^{(\La)}\big)\backslash \{ \mu_n\}\Big)>K\alpha^2.
\]
By functional calculus, we finally get the norm convergence of $(\psi_{a;n})_n$ to $\psi_a$ in $L^2$.

\noindent -- This proves that
\[
\text{s}.\,\lim_n P_n=\chi_{(-\infty,0)}\big(D_{Q_\infty}^{(\La)}\big)+\ket{\psi_a}\bra{\psi_a}-\ket{\Isym \psi_a}\bra{\Isym \psi_a}\in \mathscr{M}_{\mathscr{I}},
\]
and ends the proof.

\subsection{Existence of a minimizer for $E_{j_0,\pm}$}
We consider a family of almost minimizers $(P_{\eta_n})_n$ of type \eqref{di_almost} where $(\eta_n)_n$ is any decreasing sequence. 
We also consider the spectral decomposition \eqref{di_formtrial} of any 

\noindent$Q_n:=P_{\eta_n}-\PP$. 

For short we write $P_n:=P_{\eta_n}$ and we replace the subscript $\eta_n$ by $n$ (for instance $\psi_n:=\psi_{\eta_n}$). Moreover, we will often write $\eps$ instead of $\eps(\mathbf{t})$.

We study weak limits of $(Q_{n})_n$. We recall that $Q_n$ can be written as follows:
\begin{equation}\label{di_spec_no}
\left\{ \begin{array}{l}
N_{+;n}=\PPP N_{+;n}=\text{Proj}\,\PPh\,\psi_{\eta_n}\text{\ and\ }N_{-;n}=\Cha N_{+;n}\Cha,\\
Q_n=N_{+;n}-N_{-;n}+\g_n,\ \mathrm{Ran}\,N_{\pm;n}\cap\mathrm{Ker}\,\g_n=\{ 0\}.
\end{array}
\right.
\end{equation}
We can suppose
\[
\psi_n=\PPP a_n(r)\Phi^+_{j_0,\eps{\mathbf{t}}},\ a_n(r)\in \mathbb{S}L^2(\mathbb{R}_+,r^2dr).
\]
\begin{remark}\label{di_newton_rem}
The functions $\psi\in\mathrm{Ran}\,N_{\pm;n}$ are "almost" radial. We recall \eqref{di_form_en_trial1}, giving
\begin{equation}\label{di_radial}
\begin{array}{| l}
\forall\,x=r\om_x\in\RR,\ |\psi(x)|\le \nlp{2}{\psi}|s_n(r)|\nlp{\infty}{\Phi^{\pm}_{j_0,\pm(j_0+\tfrac{1}{2})}},\\
4|s_n(r_0)|^2:=\big|(1+\frac{g_0(|\nabla|)}{|\D|})a_n\big|(r_0)^2+\big|\frac{g_1(|\nabla|)}{|\nabla||\D|}(\partial_r a_n+\eps \tfrac{a_n}{r})\big|(r_0)^2.
\end{array}
\end{equation}
In particular by Newton's Theorem for radial function we have:
\begin{equation}\label{di_newton}
\forall\,\psi\in \mathrm{Ran}\,N_{\pm;n},\ |\psi|^2*\frac{1}{|\cdot|}(x_0)\le K(j_0)\frac{\nlp{2}{\psi}^2}{|x_0|}.
\end{equation}
\end{remark}

\noindent -- We first prove that there is no vanishing, that is
\[
\exists A>0,\ \limsup_n \sup_{z\in\RR}\underset{B(z,A)}{\dint}|\psi_{n}(x)|^2dx>0.
\]
Indeed, let assume this is false. Then using \eqref{di_newton}, it is clear that
\[
\nqq{N_{\pm;n}}^2\to 0,
\]
and we get $\liminf \mathcal{E}^0_{\text{BDF}}\ge 2(2j_0+1)m+\liminf \mathcal{E}^0_{\text{BDF}}(\g_n)\ge 2(2j_0+1)m,$

\noindent an inequality that is false as shown in the previous section.

\begin{center}\textbf{Thus, we have: $Q_n\rightharpoonup Q_{\infty}\neq 0$.} 
\end{center}

\noindent -- As the BDF energy is sequential weakly lower continuous \cite{Sc}, we have
\[
E_{j_0,\eps}\ge \mathcal{E}_{\text{BDF}}^0(Q_{\infty}).
\]
Our aim is to prove that $Q_{\infty}+\PP\in\mathscr{W}_{\mathbf{t}X^{\ell_0}}$: in other words that $Q_{\infty}$ is a minimizer for $E_{j_0,\eps}$.

The spectral decomposition \eqref{di_spec_no} is not the relevant one: let us prove we can describe $P_n$ in function of the spectral spaces of the "mean-field operator" $\wt{D}_{Q_n}$: the first step is to prove \eqref{di_spec_yes} below.

We recall that $Q_n$ satisfies Eq. \eqref{di_eq_almost}, that we have the decomposition \eqref{di_decomp_hl}.

Using \eqref{di_alm_min}, we have for all $\psi$ in $\mathbb{S}\mathrm{Ran}\,N_{+;n}$:
\begin{align*}
\psh{\wt{D}_{Q_n} \psi}{\psi}-m&=\psh{(|\D|-m)\psi}{\psi}- \psh{(\alpha R_{Q_n}+2\eta_n \G_n )\psi}{\psi},\\
       &\apprge -\alpha \nqq{Q_n}\nlp{2}{\,|\nabla|^{1/2}\psi}-\\eta_n\ns{2}{\G_n}\apprge -\alpha^2(2j_0+1).
\end{align*}
Thus $\mathrm{Ran}\,P_n\cap \mathrm{Ran}\,\pvt^{n}_+\neq \{0\}.$ 

\noindent -- Let us prove this subspace has dimension $2j_0+1$: we use the minimizing property of $Q_n$. The condition on the first derivative gives \eqref{di_eq_almost}.
The estimation of the energy (from above and below) obtained in the previous section gives this result. Indeed, using the Cauchy expansion and the method of \cite{sok}, we have
\begin{equation}\label{di_kin_no_proof}
\begin{array}{|l}
 \sqrt{\ttr\big(|\D|\g_{vac;n}^2\big)}\apprle \alpha(\nqq{Q_n}+\eta_n\ns{2}{\G_n})\apprle \sqrt{L\alpha}\sqrt{\alpha j_0} ,\\
 \g_{vac;n}:=\chi_{(-\infty,0)}\big(\wt{D}_{Q_n}\big)-\PP.
\end{array}
\end{equation}
 The Cauchy expansion is explained in \eqref{di_cauchy0}-\eqref{di_cauchy20} below, we assume the above estimate  for the moment (see \eqref{di_estim_kin}).

We write $Q_n=N_n+\ov{\g}_{n}$: there holds
\[
\big|\ns{2}{N_n}^2-\ns{2}{Q_n}^2\big|\apprle L^{1/2}\alpha(2j_0+1).
\]
As $2(2j_0+1)\le \ns{2}{Q_n}^2\le 2(2j_0+1)\big(1-\alpha \pi/4\big)^{-1}$, then necessarily
\begin{equation}\label{di_arg_en}
\big|\ns{2}{N_n}^2-2(2j_0+1)\big|\apprle \alpha (2j_0+1),
\end{equation}
and for $\alpha$ sufficiently small, the upper bound is smaller than $4$. This proves
\[
\text{Dim}\,\mathrm{Ran}\,P_n\cap \mathrm{Ran}\,\pvt^{n}_+=2j_0+1.
\]
\begin{remark}\label{di_form_nn}
There exists a unitary $\psi_{a;n}$ such that 
\[
\PPh\,\psi_{a;n}=\mathrm{Ran}\,P_n\cap \mathrm{Ran}\,\pvt^{n}_+.
\]
We can assume that $\psi_{a;n}\in\mathrm{Ker}\big( \mathrm{J}_3-j_0\big)$. Then we have
\begin{equation}
N_n:=\text{Proj}\,\PPh\,\psi_{a;n}-\text{Proj}\,\PPh\,\Cha\psi_{a;n}.
\end{equation}
Equivalently writing $\psi_{w;n}:=\Cha \psi_{a;n}$ there holds $\PPh\,\psi_{w;n}=\mathrm{Ran}\,(1-P_n)\cap \mathrm{Ran}\,\pvt^{n}_-$.
\end{remark}


\noindent -- We have:
\begin{equation}\label{di_spec_yes}
P_n=\text{Proj}\,\PPh\,\psi_{a;n}-\text{Proj}\,\PPh\,\psi_{w;n}+\pvt_-^n.
\end{equation}
We thus write
\begin{equation}
Q_n=N_n+\g_{vac;n}.
\end{equation}

As $ \mathrm{Ran}\,P_n$ is $\wt{D}_{Q_n}$ invariant and that $\wt{D}_{Q_n}$ is bounded (with a bound that depends on $\La$), necessarily
\[
\wt{D}_{Q_n}\psi_{a;n}=\mu_n\psi_{a;n},\ \mu_n\in \mathbb{R}_+.
\]
As in \cite{pos_sok}, studying the Hessian we have
\[
m-\mu_n+2\eta_n\ge 0.
\]

\noindent -- As for $\psi_n$, there is no vanishing for $(\psi_{a,n})_n$ for $\alpha$ sufficiently small: decomposing $\psi_+\in\mathrm{Ran}\,P_n$:
\[
\psi_+=a\psi_{a;n}+\phi,\ \phi\in \mathrm{Ran}\,P_n\cap \mathrm{Ran}\,\pvt_-^n,
\]
we have
\[
|a|^2\ge\frac{1}{\mu}\big(m+\psh{|\wt{D}_{Q_n}|\phi}{\phi}-K(\alpha^2j_0+\eta_n\ns{2}{\G_n}) \big).
\]
Provided that $\mu_n$ is close to $m$, the absence of vanishing for $\psi_n$ implies that of $\psi_{a;n}$.

By Kato's inequality \eqref{di_kato}:
\begin{align*}
\wt{D}_{Q_n}^2&\ge |\D|\big(1-2\alpha\nb{R_{Q_n}|\D|^{-1}}-4\eta_n \nb{\G_n}\big)|\D|\\
			&\ge |\D|^2\big(1-\alpha  \nqq{Q_n}-4\eta_n \ns{2}{\G_n}\big)
\end{align*}
Thus
\[
\big| \wt{D}_{Q_n}\big|\ge |\D|\big(1-\alpha \nqq{Q_n}-2\eta_n \ns{2}{\G_n}\big)\text{\ and\ }\mu_n\ge 1-K(\alpha^2 j_0 +\eta_n \ns{2}{\G_n}).
\]
In the same way we can prove that
\[
|\mu_n-m|\apprle \alpha^2j_0+\eta_n\ns{2}{\G_n}
\]
So
\[
\psi_{a,n}\rightharpoonup\psi_{a}\neq 0.
\]

\noindent -- We decompose $\g_{vac;n}=\pvt_-^n-\PP\in\mathscr{W}_{0}-\PP$ as in \eqref{di_formtrial}: using Cauchy's expansion \eqref{di_cauchy0}-\eqref{di_cauchy20}, we have
\begin{equation}\label{di_cauchy}
\pvt_-^n-\PP=\frac{1}{2\pi}\dint_{-\infty}^{+\infty}\frac{d \om}{\D+i\omega}\big(2\eta_n \G_n-\alpha\Pi_\La R_{Q_n}\Pi_\La+2\eta_n \G_n \big)\dfrac{1}{\wt{D}_{Q_n}+i\om}\Pi_\La.
\end{equation}
To justify this equality, we remark that $|\wt{D}_{Q_n}|$ is uniformly bounded from below, it follows that the r.h.s. of \eqref{di_cauchy} is well-defined provided that $\alpha\le \alpha_{j_0}$:
\begin{align*}
 \Pi_\La R_{Q_n}\Pi_\La^2&\apprle |\nabla|\nqq{Q_n}^2\apprle \alpha(2j_0+1)|\nabla|\le \alpha(2j_0+1)|\D|^2.
\end{align*}
We must ensure that $\alpha \sqrt{\alpha(2j_0+1)}$ is sufficiently small.

Integrating the norm of bounded operator in \eqref{di_cauchy}, we obtain
\[
\nb{\pvt_-^n-\PP}\apprle \alpha \nqq{Q_n}+\eta_n\ns{2}{\G_n}<1.
\]

We also expand in power of $Y_n:=-\alpha \Pi_\La R_{Q_n}\Pi_\La +2\eta_n \G_n$ as in \eqref{di_cauchy20}
\begin{equation}\label{di_cauchy2}
	\begin{array}{rcl}
		\pvt_-^n-\PP&=&\ssum_{j\ge 1}\alpha^j M_j[Y_n].
	\end{array}
\end{equation}
We have
\begin{equation}\label{di_estim_gn1}
\ns{2}{\g_{vac;n}}\apprle \alpha \nqq{Q_n}+\eta_n\ns{2}{\G_n}\apprle \alpha^2.
\end{equation}
We take the norm $\ns{2}{|\D|^{1/2}(\cdot)}$:
\begin{equation}\label{di_estim_kin1}
\ns{2}{|\D|^{1/2}\g_{vac;n}}\apprle \sqrt{L\alpha}\nqq{Q_N}+\eta_n\ns{2}{\G_n}\apprle L^{1/2}\alpha j_0.
\end{equation}

\noindent -- We thus write
\[
\begin{array}{rcl}
\g_{vac;n}&=&\ssum_{j\ge 1}\la_{j;n}q_{j;n},
\end{array}
\]
where $q_{j;n}$ has the same form as the one in \eqref{di_formtrial}. 

Up to a subsequence, we may assume all weak convergence as in Remark \eqref{di_diag_extrac}: the sequence of eigenvalues $(\la_{j;n})_n$ tends to $(\mu_j)_j\in\ell^2$ and each $(e_{j;n}^\star)_n$ (with $\star\in\{a,b\}$) tends to $e_{j;\infty}^\star$, $(\psi_{e;n})_n$ tends to $\psi_{e}$. We can also assume that the sequence $(\mu_n)_n$ tends to $\mu$ with $0\le \mu\le m$. 

\begin{notation}
For shot we write $\psi_v:=\Cha \psi_e$. 

Furthermore, we write $\ov{P}:=Q_{\infty}+\PP$ and $\pvac:=\chi_{(-\infty,0)}(D_{Q_\infty}^{(\La)})$. 
\end{notation}

\noindent -- We will prove that

\begin{enumerate}
	\item $\big[D^{(\La)}_{Q_\infty} ,\ov{P}\big]=0$,
	\item $D_{Q_\infty}^{(\La)}\psi_{a}=\mu\psi_a$ and so $\pvac \psi_a=0$. 
	
		Moreover $D_{Q_\infty}^{(\La)}\Cha \psi_{a}=-\mu\Cha \psi_a$ and $\psh{\Cha \psi_a}{\psi_a}=0$.
	\item 
	\begin{equation}\label{di_marre_form}
	\pvac=\ov{P}-\text{Proj}\,\PPh(\psi_a)+\text{Proj}\,\PPh(\Cha \psi_a)=:\ov{P}-N. 
	\end{equation}

\end{enumerate}

These results follow from the strong convergence
\begin{equation}\label{di_strong}
\text{s}.\,\lim_n R_{Q_n}=R_{Q_\infty}.
\end{equation}
This fact enables us to show
\begin{equation}
\begin{array}{|l}
 \lim_n R_{Q_n}\psi_{a;n}=R_{Q_\infty}\psi_a\text{\ in\ }L^2,\\
\text{s.\,op.}\ \lim_n\big(\pvt_-^n-\PP\big)=\pvac-\PP\text{\ in\ }\mathcal{B}(\hl),\\
\text{w.\,op.}\ \lim_n P_n=\pvac-\PP+\text{Proj}\,\PPh\,\psi_a-\text{Proj}\,\PPh\psi_w\text{\ in\ }\mathcal{B}(\hl),\\
\lim_n\psi_{a;n}=\psi_a\text{\ in\ }L^2.
\end{array}
\end{equation}

\begin{remark}\label{di_assume}
We only write in this paper the proof of
\[
R_{Q_n}\psi_{a;n}\underset{n\to+\infty}{\overset{L^2}{\longrightarrow}}R_{Q_\infty}\psi_a\text{\ and\ }\psi_{a;n}\underset{n\to+\infty}{\overset{L^2}{\longrightarrow}}\psi_a.
\]
The convergence in the weak-topology can be proved using the same method as in \cite{pos_sok}. For the first limit this follows from the convergence of $R_{Q_n}$ in the strong topology. For the proof of this fact and of the strong convergence of $\g_{vac;n}=\pvt_-^n-\PP$, we refer the reader to \cite{pos_sok}. 

For $R_{Q_n}$, it suffices to remark that $Q_n(x,y)$ converges in $L^2_{loc}$ and $a.e.$. To estimate the mass at infinity, we simply use the term $\tfrac{1}{|x-y|}$ in $\tfrac{Q_n(x,y)}{|x-y|}$.

The strong convergence of $\g_{vac;n}$ follows from that of $R_{Q_n}$ and the Cauchy expansion \eqref{di_cauchy2}.

Then, assuming all these convergences, the convergence of $Q_n$ resp. $\big[ \wt{D}_{Q_n}^{(\La)}; P_n\big]$ in the weak operator topology to $Q_\infty$ resp. $\big[ D_{Q_\infty}^{(\La)},\ov{P}\big]$ are straightforward.

Similarly, using \eqref{di_strong}, it is clear that
\[
\wt{D}_{Q_n}\psi_{a;n}\underset{n\to+\infty}{\rightharpoonup}D_{Q_\infty}\psi_a,
\]
and that
\[
D_{Q_\infty}^{(\La)}\psi_a=\mu\psi_a.
\]
To get the existence of minimizer, it suffices to prove that $\nlp{2}{\psi_a}=1$ or equivalently $\lim_n\psi_{a;n}=\psi_a$ in $L^2$.
\end{remark}

\medskip

\noindent -- To prove the norm convergence of $\psi_{a;n}$ to $\psi_a$, we need a uniform upper bound of $\mu_n$, or precisely, we need the following:
\begin{equation}\label{di_need}
\limsup_n(m-\mu_n)>0.
\end{equation}
Indeed, we then get
\begin{equation}\label{di_lim_l2}
(\D-\mu_n)\psi_{a;n}=\alpha R_{Q_n}\psi_{a;n}-2\eta_n \G_n\psi_{a;n}\text{\ and\ }\psi_{a;n}=\frac{\alpha}{\D-\mu_n}\big(R_{Q_n}\psi_{a;n}-2\eta_n \G_n\psi_{a;n}\big).
\end{equation}
Provided that \eqref{di_need} holds and that we have norm convergence of $R_{Q_n}\psi_{a;n}$ we obtain the norm convergence of $\psi_{a;n}$.

\noindent -- To prove the norm convergence of $R_{Q_n}\psi_{a;n}$ to $R_{Q_\infty}\psi_a$, we use the fact that the element of $\PPh\,\psi_{a;n}$ are "almost radial" (see in Remark \ref{di_newton_rem}). We recall  \eqref{di_newton} holds. In the following, we write $\delta Q_n:=Q_n-Q_\infty$ and $\delta \psi_n:=\psi_{a;n}-\psi_a$ and use Cauchy-Schwartz inequality: for any $A>0$ there hold
\begin{align*}
\dint_{|x|\ge A}\Big|\dint \frac{\delta Q_n(x,y)}{|x-y|}\psi_{a;n}(y)dy\Big|^2dx&\le \nqq{\delta Q_n}^2 \frac{K(j_0)}{A},\\
\dint_{|x|\le A}\Big|\dint \frac{\delta Q_n(x,y)}{|x-y|}\psi_{a;n}(y)dy\Big|^2dx&\le \frac{2\pi}{2}\psh{|\nabla| \psi_{a;n}}{\psi_{a;n}}\underset{B(0,A)\times B(0,2A)}{\diint}\frac{|\delta Q_n(x,y)|^2}{|x-y|}dxdy\\
 &\ \ \ +\frac{2}{A^2}\ns{2}{\delta Q_n}^2\nlp{2}{\psi_{a;n}}^2.
\end{align*}
Thus
\[
\limsup_n \nlp{2}{R[Q_n-Q_\infty]\psi_{a;n}}=0.
\]
Similarly
\begin{align*}
\dint_{|x|\ge A}\Big|\frac{Q_\infty(x,y)}{|x-y|}\delta \psi_n(y)dy\Big|^2dx&\le \frac{2}{A-\tfrac{A}{2}}\ns{2}{Q_\infty}^2\nlp{2}{\delta\psi_n}^2+2\nlp{2}{\delta \psi_n}^2\frac{2}{A}\nqq{Q_\infty}^2,\\
\dint_{|x|\le A}\Big|\frac{Q_\infty(x,y)}{|x-y|}\delta \psi_n(y)dy\Big|^2dx&\le\frac{2\pi}{2}\psh{|\nabla|\delta \psi_n}{\delta\psi_n}\underset{B(0,A)\times B(0,2A)}{\diint}\frac{|\delta Q_n(x,y)|^2}{|x-y|}dxdy\\
&\ \ \ +\frac{2}{A^2}\ns{2}{Q_\infty}^2\nlp{2}{\delta\psi_n}^2,
\end{align*}
and
\[
\limsup_n\nlp{2}{R_{Q_\infty}(\psi_{a;n}-\psi_a)}=0.
\]
This proves that
\[
\lim_{n\to+\infty}\nlp{2}{R_{Q_n}\psi_{a;n}-R_{Q_\infty}\psi_a}=0.
\]

\noindent -- Let us prove \eqref{di_need}. We have:
\begin{equation}\label{di_need_to_proof}
\begin{array}{rcl}
2\mu_n(2j_0+1)&=&\ttr\Big(\wt{D}_{Q_n}N_n\Big),\\
 &=&\ttr\Big(\wt{D}_{\g_{vac;n}}N_n\Big)-\alpha \nqq{N_n}^2,\\
 &=&\mathcal{E}^0_{\text{BDF}}(Q_n)-\mathcal{E}^0_{\text{BDF}}(\g_{vac;n})-\frac{\alpha}{2}\nqq{N_n}^2,\\
 &<&2m(2j_0+1)-K(j_0)\alpha^2.
\end{array}
\end{equation}
This upper bound holds provided that $\alpha\le \alpha_{j_0}$ thanks to the upper bound of $E_{j_0,\eps}$ obtained in the previous section.

\subsection{Lower bound of $E_{j_0,\pm}$}\label{di_low_bound}
Our aim is to prove the estimate of Proposition \ref{di_est}. We consider the minimizer $Q_{\infty}=N+\g_{vac}$ found in the previous subsection. It satisfies Eq. \eqref{di_marre_form} where 
\begin{equation}\label{di_marre_form_re}
\ov{P}=\PP+Q_\infty\text{\ and\ }\g_{vac}=\chi_{(-\infty,0)}(D_{Q_infty}^{(\La)})-\PP.
\end{equation}

\noindent -- The proof is the same as that in \cite{sok,pos_sok} and relies on estimates on the Sobolev norms $\ns{2}{\,|\nabla|^s N_+}$ where we write
\begin{equation}\label{di_eq_recall}
 N_+:=\text{Proj}\,\PPh\,\psi_a=\mathrm{Ker}\,(D_{Q_\infty}^{(\La)}-\mu).
\end{equation}
Using \eqref{di_eq_recall}, we get
\begin{align*}
 \ttr\big(|\D|^2N_+\big)&= 2(2j_0+1)\mu^2+2\alpha \mu\ttr\big(R_{Q_\infty}N_+\big)+\alpha^2\ttr\big(R_{Q_\infty}^2N_+\big),\\
                       &\le 2(2j_0+1)\mu^2+4\alpha \mu\ns{2}{Q_\infty}\ns{2}{\nabla N_+}+4\alpha^2\ns{2}{Q_\infty}^2\ns{2}{\nabla N_+}^2
\end{align*}
and provided that $\alpha\le \alpha_{j_0}$, we get
\begin{equation}
 \ttr\big((-\Delta)N_+\big)\apprle \frac{\alpha^2(2j_0+1)}{1-4\alpha^2(2j_0+1)-2\nlp{\infty}{g_0}\nlp{\infty}{g_0''}}.
\end{equation}
We have used Hardy's inequality: 
\begin{equation}\label{di_hardy}
\dfrac{1}{4|\cdot|^2}\le -\Delta\text{\ in\ }\RR.
\end{equation}
We recall that
\[
0\le \nlp{\infty}{g_0}-1\apprle \alpha\llo\text{\ and\ } \nlp{\infty}{g_0''}\apprle \alpha.
\]
See \eqref{di_estim_g} (or \cite[Appendix A]{sok} for more details).

Thus for sufficiently small $\alpha$, we have
\begin{equation}\label{di_wf_to_scale}
 \forall\,\psi\in\mathbb{S}\mathrm{Ran}\,N_+,\ \nlp{2}{\nabla \psi}^2\apprle \frac{\alpha^2}{1-4\alpha^2(2j_0+1)-2\nlp{\infty}{g_0}\nlp{\infty}{g_0''}}\apprle \alpha^2.
\end{equation}

\noindent -- By \emph{bootstrap argument}, we can estimate $\ns{2}{\,\Delta N_+}$. We have:
\begin{equation}\label{di_marre_boot}
 \forall\,\psi\in\mathbb{S}\mathrm{Ran}\,N_+,\ \nlp{2}{\,|\nabla|^{3/2}\psi}^2\apprle \alpha^{3}\sqrt{2j_0+1}\text{\ and\ }\nlp{2}{\Delta \psi}\apprle \alpha^4(2j_0+1)^{3/2}.
\end{equation}
We prove this result below.

Furthermore, using the Cauchy expansion \eqref{di_cauchy20} and \eqref{di_cauchy_est0}, we get
\[
 \begin{array}{| rcl}
  \ns{2}{\,|\D|^{1/2}\g_{vac}}&\apprle&\alpha \ns{2}{\nabla N}+\sqrt{L\alpha}\nqq{\g_{vac}}+\alpha^2\nqq{Q_\infty}^2\big(\ns{2}{\nabla N}+\nqq{\g_{vac}} \big),
 \end{array}
\]
hence
\begin{equation}\label{di_marre_gvac}
 \ns{2}{\,|\D|^{1/2}\g_{vac}}\apprle \alpha^2\sqrt{2j_0+1}.
\end{equation}

Now, if we assume \eqref{di_marre_boot}-\eqref {di_marre_gvac}, then we get 
\[
\text{For\ } \alpha\le \alpha_{j_0},\ \mathcal{E}^0_{\text{BDF}}\big(Q_\infty\big)=2m(2j_0+1)+\frac{\alpha^2m}{g'_1(0)^2}E_{\mathbf{t}X^{\ell_0}}^{nr}+\mathcal{O}\big(\alpha^3 K(j_0)\big).
\]
We do not prove this fact: the method is the same as in \cite{sok,pos_sok} (in the proof of the lower bound of $E^0_{\text{BDF}}(1)$ resp. $E_{1,1}$).

We just recall how we get \eqref{di_marre_boot}.

\subparagraph{Proof of \eqref{di_marre_boot}}
We scale the wave functions of \eqref{di_wf_to_scale} by $\la:=\frac{g'_1(0)^2}{\alpha m}$:
\[
 \forall\,x\in \RR,\ U_{\la}\psi(x)=\un{\psi}(x):=\la^{3/2}\psi(\la x),
\]
and we split $\psi$ (resp. $\un{\psi}$) into the upper spinor $\ph$ (resp. $\un{\ph}$) and the lower spinor $\chi$ (resp. $\un{\chi}$).
Thanks to \eqref{di_need_to_proof}, we have
\[
 \alpha^{-2}(m-\mu)=:\alpha^{-2}\delta m\ge K(j_0)>0
\]
provided that $\alpha$ is sufficiently small $(\alpha\le \alpha_{j_0})$.

We write
\[
 \forall\,Q_0\in\mathfrak{S}_2,\ \un{Q_0}:=U_{\la} Q_{0}U_{\la}^{-1}=U_{\la} Q_{0}U_{\la^{-1}}.
\]
For all $\psi$ in $\mathbb{S}\mathrm{Ran}\un{N_+}$ we have
\begin{equation}\label{di_eq_scale}
 \left\{\begin{array}{rcl}
          \la^2\delta m\un{\ph}&=&i\la \boldsymbol{\sigma}\cdot \nabla \un{\chi}+\alpha \la \big(R_{\un{Q}_\infty} \un{\psi}\big)_{\uparrow},\\
          \un{\chi}&=&\frac{-i\la\boldsymbol{\sigma}\cdot \nabla \un{\ph}}{\la(m+\mu)}-\tfrac{\alpha}{\la} \big(R_{\un{Q}_\infty} \un{\psi}\big)_{\downarrow}.
        \end{array}
\right.
\end{equation}
\noindent -- We recall
\begin{equation}\label{di_marre_Rtrois}
\forall\,Q_0\in\mathfrak{S}_2,\  \lVert \big[\nabla,R_{Q_0}\big]\tfrac{1}{|\nabla|^{1/2}} \rVert_{\mathcal{B}}^2\apprle \diint |p-q|^2|p+q||\wh{Q_0}(p,q)|^2dpdq.
\end{equation}
This result was previously proved in \cite{pos_sok} and follows from the fact that a (scalar) Fourier multiplier $F(\mathbf{p}-\mathbf{q})=F(-i\nabla_x+i\nabla_y)$ commutes with the operator $R[\cdot]:Q(x,y)\mapsto \tfrac{Q(x,y)}{|x-y|}$. Then it suffices to use Hardy's inequality \eqref{di_hardy}:
\[
 \nlp{2}{\big[\nabla,R_{\un{Q_\infty}}\big]\un{\psi}}^2\apprle \la^2\diint |p-q|^2|\wh{Q}_{\infty}(p,q)|^2dpdq\times \nlp{2}{\nabla \un{\psi}}^2.
\]

By Hardy's inequality \eqref{di_hardy} and \eqref{di_marre_Rtrois}, the following holds:
\begin{equation}\label{di_est_un_scale}
 \begin{array}{| rcl}
  \ns{2}{\un{\chi}}^2&\le& \frac{2}{4\la^2m^2}\ns{2}{\nabla \un{\ph}}^2+2\alpha^2\ns{2}{R_{\un{Q_\infty}}\un{\psi}}^2\apprle \alpha^2,\\
  \ns{2}{\nabla \un{\chi}}^2&\le&2(\la \delta m)^2+2\alpha^2\ns{2}{R_{\un{Q_\infty}} \un{\psi}}^2\apprle \frac{(\delta m)^2}{\alpha^2}+\alpha^2(2j_0+1),\\
  \ns{2}{\Delta \un{\ph}}^2&\le&2\la^2 m\nlp{2}{\nabla \un{\chi}}^2+2\alpha^2(\nlp{2}{\big[\nabla,R_{\un{Q_\infty}}\big]\un{\psi}}+\nlp{2}{R_{Q_\infty}}\nabla \un{\psi})^2  \\
                        &\apprle& \frac{(\delta m)^2}{\alpha^4}+(2j_0+1)+\alpha^2(2j_0+1)^{3/2},  \\
  \ns{2}{\Delta \un{\chi}}^2&\le& 2\la^2(\delta m)^2\nlp{2}{\nabla \un{\ph}}+2\alpha^2(\nlp{2}{\big[\nabla,R_{\un{Q_\infty}}\big]\un{\psi}}+\nlp{2}{R_{Q_\infty}}\nabla \un{\psi})^2  \\
			&\apprle& \frac{(\delta m)^2}{\alpha^2}+(2j_0+1)+\alpha^2(2j_0+1)^{3/2}. 
 \end{array}
\end{equation}

\noindent -- There remains to estimate
\[
 \diint |p-q|^2|\wh{Q_0}(p,q)|^2dpdq,\ \text{for\ }Q_0=N\text{\ and\ }\g_{vac}.
\]
For $Q_0=N$, we just have to estimate $\ttr\big(|\nabla|^2 N_+\big)$. 

The case $Q_0=\g_{vac}$ is dealt with as in \cite{sok,sokd}: by a \emph{fixed-point} argument (valid for $\alpha\le \alpha_{j_0}$), we prove that
\[
 \left\{\diint |p-q|^2|\wh{\g_{vac}}(p,q)|^2dpdq\right\}^{1/2}\apprle \alpha\min\big(\ns{2}{\Delta N},\ns{2}{\,|\nabla|^{3/2}N}\big).
\]

Now, we can prove that
\[
 \ttr\big(|\nabla|^{3}N_+\big)\apprle \alpha^{5/2}(2j_0+1)^{3/2}.
\]

For a unitary $\psi$ in $\mathrm{Ran}\,N_+$, there holds
\begin{equation}\label{di_sobtrois}
\begin{array}{rcl}
 \nlp{2}{\,|\nabla|^{1/2}\D\psi}^2&\le& \mu^2\psh{|\nabla|\psi}{\psi}+\alpha K \nlp{2}{\,|\nabla|^{1/2}\psi}\nlp{2}{R_{Q_\infty}\psi}\\
                                  &&\ \ \ +\alpha^2\big(\nlp{2}{[R_{Q_\infty},|\nabla|^{1/2}] \psi}+2\ns{2}{Q_\infty}\nlp{2}{\,|\nabla|^{3/2}}\big)^2.
\end{array}
\end{equation}
Similarly, in Fourier space we have:
\[
 \Big|\mathscr{F}\big([R_{Q_\infty},|\nabla|^{1/2}];p,q\big) \Big|\apprle |p-q|^{1/2}|\wh{R}_{Q_\infty}(p,q)|,
\]
and by Hardy's inequality
\[
 \nlp{2}{[R_{Q_\infty},|\nabla|^{1/2}] \psi}^2\apprle \diint |p-q| |\wh{Q_\infty}(p,q)|^2dpdq\nlp{2}{\nabla \psi}^2\apprle \ttr\big(|\nabla| Q_\infty^2\big)\nlp{2}{\nabla \psi}^2.
\]
Substituting in \eqref{di_sobtrois}, we get
\[
 \psh{|\nabla|^3\psi}{\psi}\apprle \alpha^{5/2}\sqrt{2j_0+1},\text{\ hence\ }\ttr\big(|\nabla|^3N_+ \big)\apprle \alpha^{5/2}(2j_0+1)^{3/2}.
\]


\subsection{Proof of Lemmas \ref{di_infimum_1} and \ref{di_non_triv}}\label{di_fait_ch}

\subsubsection{Proof of Lemma \ref{di_infimum_1}}

We consider a trial state $P_\psi\in\mathscr{M}_{\mathscr{I}}^1$:
\[
Q_\psi:=P_\psi-\PP=\ket{\psi}\bra{\psi}-\ket{\Isym\psi}\bra{\Isym\psi},\ \PPP\psi=\psi\in\mathbb{S}\,\hl.
\]
Its BDF energy is
\begin{align*}
\mathcal{E}^0_{\text{BDF}}(Q_\psi)&=2\psh{|\D|\psi}{\psi}-\frac{\alpha}{2}\diint\frac{|\psi\wedge \Isym\psi(x,y)|^2}{|x-y|}dxdy\\
                       &\ge 2m+2\psh{\big(|\D|-m\big)\psi}{\psi}-\alpha D\big(|\psi|^2,\psi^2\big)=:2m+\mathcal{G}_{\Isym}(\psi).
\end{align*}
We recall the following
\begin{align*}
|\D|-m&=\frac{1}{|\D|+m}\big((g_0(-i\nabla)-m)(g_0(-i\nabla)+m)+g_1(-i\nabla)^2\big).
\end{align*}

Thanks to Estimates \eqref{di_estim_g} and Kato's inequality \eqref{di_kato}, we have
\[
\mathcal{G}_{\Isym}(\psi)\le (1-K\alpha)\psh{\tfrac{-\Delta}{2|\D|} \psi}{\psi}-\alpha\frac{\pi}{4}\psh{|\nabla|\psi}{\psi}
\]
We split $\psi$ into two with respect to the frequency cut-off $\Pi_{\alpha K_0}$: we get
\[
\psi=\Pi_{\alpha K_0}\psi+\psi_2=\psi_1+\psi_2.
\]
The constant $K_0$ is chosen such that
\[
\frac{\alpha^2 K_0^2}{2\ed{\alpha K_0}}\apprge \alpha \pi \alpha K_0.
\]
Then we have
\begin{align*}
D\big(|\psi|^2,|\psi|^2\big)&=D\big(|\psi_1|^2,|\psi_1|^2\big)+\mathcal{O}\big(\psh{|\nabla|\psi_2}{\psi_2}+\ncc{|\psi_1|^2}\nlp{2}{\,|\nabla|^{1/2}\psi_2}\big)\\
 &=D\big(|\psi_1|^2,|\psi_1|^2\big)+\mathcal{O}\big(\psh{|\nabla|\psi_2}{\psi_2}+\sqrt{\alpha}\nlp{2}{\,|\nabla|^{1/2}\psi_2}\big),
\end{align*}
where we recall that $\ncc{\rho}^2=D(\rho,\rho)$. This gives
\begin{equation}
\begin{array}{rcl}
\tfrac{1}{2}\mathcal{G}_{\Isym}(\psi)&=&\psh{\frac{g_1(-i\nabla)^2}{|\D|+m}\psi_1}{\psi_1}-\alpha\frac{\pi}{2}D\big(|\psi_1|^2,|\psi_1|^2\big)\\
        &&\ \ \ +K\psh{\frac{g_1^2(-i\nabla)}{|\D|} \psi_2}{\psi_2}+\mathcal{O}(\alpha^3),\\
        &\ge&\frac{\alpha^2 g'_1(0)^2}{2m}\nlp{2}{\nabla\psi_1}^2-\frac{\alpha}{2}D\big(|\psi_1|^2,|\psi_1|^2\big)+\mathcal{O}(\alpha^3),\\
        &\ge&\frac{\alpha^2 m}{2g'_1(0)^2}E_{\text{PT}}(1)+\mathcal{O}(\alpha^3).
\end{array}
\end{equation}

\noindent We have obtained a lower bound. Let us prove that it is attained up to an error $\mathcal{O}(\alpha^3)$. That is let us prove there exists a unitary $\psi_0\in\mathrm{Ran}\PPP$ such that
\begin{equation}\label{di_di_test}
\begin{array}{rcl}
\mathcal{E}^0_{\text{BDF}}(Q_{\psi_0})-2m&=&\mathcal{G}_{\Isym}(\psi_0)+\mathcal{O}(\alpha^3)\\
                 &=& \frac{\alpha^2 m}{g'_1(0)^2}E_{\text{PT}}(1)+\mathcal{O}(\alpha^3).
\end{array}
\end{equation}

As in \cite{pos_sok}, we consider the unique positive radially symetric Pekar minimizer  $\phi_{\text{PT}}$ in $L^2(\RR,\mathbb{C})$. We form
\begin{equation}\label{di_test0}
\phi_1:=\begin{pmatrix}\phi_{\text{PT}}\\ 0\\0\\0 \end{pmatrix}\in L^2(\RR,\CC),
\end{equation}
which is a Pekar minimizer in the space of spinors. We scale this wave function by $\la^{-1}:=\frac{\alpha m}{g'_1(0)^2}$:
\begin{equation}\label{di_test1}
\forall\,x\in\RR,\ \phi_{\la^{-1}}(x):=\la^{-3/2}\phi_1(\la^{-1}x).
\end{equation}
To get a proper $\psi_0\in\mathrm{Ran}\,\PPP$, we form
\begin{equation}\label{di_test2}
\psi_0:=\frac{1}{\nlp{2}{\PPP \phi_{\la^{-1}}}}\PPP \phi_{\la^{-1}}.
\end{equation}
Our trial state is:

\begin{equation}\label{di_test3}
Q_0:=\ket{\psi_0}\bra{\psi_0}-\ket{\Isym\psi_0}\bra{\Isym \psi_0}.
\end{equation}

We do not compute its energy: the method is as in \cite{pos_sok} (except that instead of $\Isym$, the operator $\Cha$ is considered in \cite{pos_sok}, but that does not change anything). Eventually we refer the reader to the proof of the upper bound of $E_{\mathbf{t}X^{\ell_0}}$ above in Section \ref{di_subscritic} for the ideas.


\subsubsection{Proof of Lemma \ref{di_non_triv}}

We remark the following fact.

\begin{lemma}\label{di_lem_orientation}

Let $\mathbb{S}_{\Isym}\subset \hl$ be the set
\[
\mathbb{S}_{\Isym}=\big\{ f\in \hl,\ \nlp{2}{f}=1,\ \psh{f}{\Isym f}=0\big\}=\big\{ f\in \hl,\ \nlp{2}{f}=1,\ \mathfrak{Im}\psh{\PP f}{\Isym \PPP f}=0\big\}.
\]

There exists a smooth angle operator  $\mathcal{A}:\mathbb{S}_{\Isym}\to \mathbb{R}/ \pi \mathbb{Z}$.

For two $\mathbb{C}$-colinear wave functions $f_1,f_2$ in $\mathbb{S}_{\Isym}$ we have $\mathcal{A}(f_1)=\mathcal{A}(f_2)$.

Furthermore we have $\mathcal{A}^{-1}(0)=\mathrm{Ran}\,\PP$ and $\mathcal{A}^{-1}(\tfrac{\pi}{2})=\mathrm{Ran}\,\PPP$.

\end{lemma}

\paragraph{Proof:} 
Let $f$ be in $\mathbb{S}_{\Isym}$: the space $\text{Span}_{\mathbb{C}}(f,\Isym f)$ is spanned by the eigenvectors $g_{-}:=\tfrac{f+i\Isym f}{\nlp{2}{f+i\Isym f}}$ and $g_+:=\tfrac{f-i\Isym f}{\nlp{2}{f-i\Isym f}}$. We have
\[
\text{Span}_{\mathbb{C}}(f,\Isym f)=\text{Span}(\PP g_{\pm},\PPP g_{\pm}).
\]
It follows that $\mathcal{P}^0_{\pm} f\parallel \mathcal{P}^0_{\pm} g_+$ and $\PP f \parallel \Isym \PPP f$. As $f\in \mathbb{S}_{\Isym}$, for $\eps\in\{ +,-\}$ with $\mathcal{P}^0_{\eps} f\neq 0$, we have
\[
\mathcal{P}^0_{-\eps} f\in \text{Span}_{\mathbb{R}}(\mathcal{P}^0_{\eps} f).
\]
Thus we have with 
\begin{equation}\label{di_di_cond}
\mathrm{Span}_{\mathbb{R}}(f,\Isym f)=\mathrm{Span}_{\mathbb{R}}(e_-,\Isym e_-),\  e_-\in\text{Ran}\,\PP\text{\ and\ }\nlp{2}{e_-}=1.
\end{equation}
 Indeed if $\PP f\neq 0$ we can choose $e_-:=\tfrac{\PP f}{\nlp{2}{\PP f}}$, else we can choose $e_-:=\Isym \tfrac{\PPP f}{\nlp{2}{\PPP f}}$.

Then we decompose $f$ w.r.t. the basis $(e_-,\Isym e_-)$ and there exists $\theta\in \mathbb{R}/(2\pi \mathbb{Z})$ with $f=\cos(\theta) e_-+\sin(\theta) \Isym e_-$. In fact the function $f\mapsto (e_-,\Isym e_-)$ that maps $f$ to a basis \eqref{di_di_cond} is bi-valued: if $(e_-,\Isym e_-)$ is a possibility, then $(-e_-,-\Isym e_-)$ is another possibility. It follows that the angle $\theta$ is defined up to $\pi$: we thus obtain a function
\[
 \mathcal{A}:\mathbb{S}_{\Isym}\to \mathbb{R}/ \pi \mathbb{Z}.
\]
The smoothness of $\mathcal{A}$ is straightforward. The end of the proof is also clear.

\hfill{\small$\Box$}

\medskip

We use the angle operator to get a mountain pass argument: see Lemma \ref{di_mount} below.

We use and Theorem \ref{di_structure} and Proposition \ref{di_chasym}.

Let $\mathscr{U}\subset \mathscr{M}_{\mathscr{I}}$ be the \emph{open} subset
\[
\mathscr{U}\subset \mathscr{M}_{\mathscr{I}}:=\Big\{P=Q+\PP\in \mathscr{M}_{\mathscr{I}},\ \text{dim}\,\mathrm{Ker}(Q-\nb{Q})=1 \Big\}.
\]
For all $P=Q+\PP\in \mathscr{U}$, the eigenspace $\mathrm{Ker}(Q-\nb{Q})$ is spanned by a unitary vector $f_0$. By $\Isym$-symmetry, we have
\[
\Isym \mathrm{Ker}(Q-\nb{Q})=\mathrm{Ker}(Q+\nb{Q}),
\]
and we have $\psh{f_0}{\Isym f_0}=0.$ By Proposition \ref{di_chasym}, the plane $\text{Span}_{\mathbb{C}}\,(f,\Isym f)$ is spanned by $f_-\in\text{Ran}\,P$ and $f_+\in\text{Ran}\,(1-P)$.

By $\Isym$-symmetry, we have $\Isym f_-\in\mathbb{R}f_+$. In other words:

\begin{center}
\textbf{the wave function $f_-$ is in $\mathbb{S}_{\Isym}$}.
\end{center}
\begin{defi}
Let $Q+\PP\in \mathscr{U}\subset \mathscr{M}_{\mathscr{I}}$ and $f_-$ as above.
We define the smooth function $\mathcal{A}_U$ as follows:
\[
\mathcal{A}_U:Q+\PP\in \mathscr{U}\subset \mathscr{M}_{\mathscr{I}}\mapsto \mathcal{A}(f_-).
\]
It is clear it does not depend on the choice of $f_-$ but is a function of $\mathbb{C}f_-$. Furthermore, we have
\[
\forall\,P\in \mathscr{U},\ \nabla \mathcal{A}_U(P)\neq 0
\]
\end{defi}

The following Lemma is an application of classical results in geometry.

\begin{lemma}\label{di_mount}
Let $\mathscr{M}_{U,\Isym}$ be the subset
\[
\mathscr{M}_{U,\Isym}:=\big\{Q+\PP\in\mathscr{U},\ \nb{Q}=1\big\}=\mathcal{A}_U^{-1}\big(\big\{\frac{\pi}{2}\big\}\big),
\]
in other words the set of projectors in $\mathscr{U}$ whose range intersects nontrivially $\mathrm{Ran}\,\PPP$. 
For any differentiable function $c:(-\eps,\eps)\to \mathscr{M}_{\mathscr{I}}$ such that $\eps>0$, $c(0)\in \mathscr{M}_{U,\Isym}$ and
\[
\ttr\big(\nabla \mathcal{A}_U(c(0))^* \frac{d}{ds}c(0)\big)\neq 0,
\]
the following holds: any sufficiently small smooth perturbation
\[
c+\delta c:(-\eps,\eps)\to \mathscr{M}_{\mathscr{I}},
\]
in the norm 
\[
\lVert  \widetilde{c}\rVert:=\sup_{s\in (-\eps,\eps)}\ns{2}{\widetilde{c}(s)-\PP}+\sup_{s\in (-\eps,\eps)}\ns{2}{\tfrac{d}{ds}\widetilde{c}(s)}
\]
still intersects $\mathscr{M}_{U,\Isym}$ at some $s(\delta c)$.
\end{lemma}

\noindent -- Let us now prove Lemma \ref{di_non_triv}. We recall that we have defined a loop $c_\psi=c_0$ that crosses $\mathscr{M}_{U,\Isym}$ at $s=\tfrac{1}{2}$ and we can easily check that $\ttr\big(\mathcal{A}_U(c(2^{-1}))^*\frac{d}{ds}c(2^{-1})\big)=1\neq 0.$

\medskip

Furthermore we have defined the family $(c_t)_{t\ge 0}$ by $c_t:=\Phi_{\text{BDF};t}(c_\psi)$ where $\Phi_{\text{BDF};t}$ is the gradient flow of the BDF energy.

\medskip

\noindent -- By Lemma \ref{di_mount}, the loop $c_t$ still intersects $\mathscr{M}_{U,\Isym}$ for sufficiently small $t$. We must ensure that this fact holds for all $t\ge 0$ to end the proof. 

We use a continuation principle and set
\[
t_{\infty}:=\sup\Big\{t\ge 0,\ \forall\,0\le \tau\le t,\exists s_0\in[0,1] c_\tau\text{\ crosses\ }\mathscr{M}_{U,\Isym}\text{\ at\ }s=s_0\Big\}.
\]
We also define for all $0\le \tau<t_{\infty}$:
\[
\begin{array}{| rcl}
 s_{-}(\tau)&=&\sup\{ s\in[0,1],\ \forall\,s'\le s,\ \nb{c_\tau(s')}<1\}>0,\\
 s_{+}(\tau)&=&\inf\{ s\in[0,1],\ \forall\,s'\ge s,\ \nb{c_\tau(s')}<1\}<1.
 \end{array}
\]
\noindent -- We assume that $t_\infty<+\infty$ and prove this implies a contradiction.

The initial loop $c_0$ induces
\[
\mathcal{L}_0:s\in [0,1]\mapsto \mathcal{A}_U(c_0(s))=\pi s\in \mathbb{T},
\]
and we notice that $\mathcal{L}_0$ has a non-trivial homotopy.

Thus, at least for $\tau$ close to $0$, the following holds.
\begin{enumerate}
 \item There exist $0<\eta_\tau,\eta_\tau'\ll 1$ such that
 \begin{equation}\label{di_cont_0}
  \mathcal{A}_U\big[c_\tau\big((s_-(\tau)-\eta_\tau,s_-(\tau)) \big)\big]\cap (\tfrac{\pi}{2},\tfrac{\pi}{2}+\eta_\tau')=\varnothing.
 \end{equation}
\item There exist $0<\eta_\tau,\eta_\tau'\ll 1$ such that
 \begin{equation}\label{di_cont_1}
  \mathcal{A}_U\big[c_\tau\big((s_+(\tau),s_+(\tau)+\eta_\tau) \big)\big]\cap (\tfrac{\pi}{2}-\eta_\tau',\tfrac{\pi}{2})=\varnothing.
 \end{equation}
\end{enumerate}

\noindent The functions $\tau\ge 0\mapsto s_{\pm}(\tau)$ are well-defined and continuous in a neighbourhood of $0$ with $ s_-(0)=s_+(0)=\tfrac{1}{2}.$

\medskip

\noindent -- We prove that by continuity in $\tau$ we have
\begin{equation}\label{di_contz}
\forall\,s\in[0,1],\ \nb{c_{\tau}(s)}=1\Rightarrow c_{\tau}(s)+\PP\in \mathscr{M}_{U,\Isym}
\end{equation}
and in particular
\begin{equation}\label{di_cont}
 c_\tau(s_\pm(\tau))\in \mathscr{M}_{U,\Isym}-\PP.
\end{equation}
If not, this implies that as $\tau$ increases, the second highest eigenvalue of $c_\tau(s_0)$ also increases to reach $1$ where \eqref{di_cont_0} becomes false, at some $(\tau_0,s_0)$.

This cannot occurs because of the energy condition: if this was true, we would have by Kato's inequality \eqref{di_kato}
\[
 \mathcal{E}^0_{\text{BDF}}\big(c_{\tau_0}(s_0)\big)\ge (1-\alpha \tfrac{\pi}{4})\ttr\big(|\D|c_{\tau_0}(s_0)^2 \big)\ge 4m(1-\alpha \tfrac{\pi}{4})>2m.
\]

Thus \eqref{di_contz}-\eqref{di_cont} hold for all $0\le \tau<t_\infty$.

\medskip

\noindent -- Thanks to this fact, by continuity for all $0\le \tau<t_\infty$, \eqref{di_cont_0}-\eqref{di_cont_1} hold: if we follow the point $s_{\pm}(\tau)$ from $\tau=0$, we see that there cannot exist $\tau_0$ such that \eqref{di_cont_0} or \eqref{di_cont_1} becomes false, because the set  $\{t\ge 0,\ \forall\,0\le \tau< t,$ \eqref{di_cont_0} (resp. \eqref{di_cont_1}) holds for $\tau \}$
is non-empty and open. 

\noindent -- Up to an isomorphism of $[0,1]$, we can suppose that for all $0\le \tau\le t_\infty$,
\[
\forall\, s\in[0,1],\ \ns{2}{\partial_s c_\tau(s_0)}\apprle 1.
\]
\begin{remark}
 In $\mathfrak{S}_2$, the function $\partial_sc_t(s_0)$ satsifies the following equation:
 \[
  \frac{d}{dt}\partial_sc_t(s_0)=\partial_s \nabla \mathcal{E}^0_{\text{BDF}}(c_t(s_0))\in\mathfrak{S}_2.
 \]
\end{remark}

These new loops are written $\wt{c}_\tau$ and have the same range as the $c_\tau$'s and define the same arc length. 

Studying the limit of $\wt{c}_\tau$ as $\tau$ tends to $t_\infty$, we get that at $t=t_\infty$, \eqref{di_cont_0}-\eqref{di_cont_1} still holds for the loop $\wt{c}_{t_\infty}$ at some $0< s_{-}(t_\infty)\le s_+(t_\infty)< 1$.

Then necessarily, the loop $\wt{c}_{t_\infty}$ crosses $\mathscr{M}_{U,\Isym}$ at some $s\in [s_-(t_\infty),s_+(t_\infty)]$. Going back to $c_{t_\infty}$, this proves that the same holds for $c_{t_\infty}$, which contradicts the definition of $t_\infty$.



\section{Proofs on results on the variational set}\label{di_proofmanif}

\subsection{Proof of Lemma \ref{di_irreduc}}\label{di_irr_proof}

Let
\[
\PPh':\mathbf{SU}(2)\to \mathbf{U}(E),\ E\subset\hl
\]
be an irreducible representation of $\PPh$. As $\Jbf^2$ and $\Spbf$ commutes with the action of $\mathbf{SU}(2)$, then necessarily $E$ is an eigenspace for $\Jbf^2$ and $\Spbf$, associated to $j(j+1)$ and $\kappa_j=\eps(j+\tfrac{1}{2})$ where $j\in \tfrac{1}{2}+\mathbb{Z}_+$ and $\eps=\pm$. The eigenspaces are known \cite[p. 126]{Th}: they are spanned by wave functions of type
\begin{equation}\label{di_irr_0}
\forall\,x=r\om_x\in\RR,\ \psi(x):=a(r)\Phi^{\pm}_{m,\kappa_j},\ m=-j,-j+1,\ldots,j,
\end{equation}
where
\begin{subequations}\label{di_courage}
\begin{equation}\label{di_irr_1}
a(r)\in L^2(\mathbb{R}_+,r^2dr),
\end{equation}
\begin{equation}\label{di_irr_2}
\Phi^+_{m,\pm(j+\tfrac{1}{2})}:=\begin{pmatrix}i\Psi^{m}_{j\pm\tfrac{1}{2}}\\ 0\end{pmatrix}\text{\ and\ }\Phi^-_{m,\pm(j+\tfrac{1}{2})}:=\begin{pmatrix}0 \\ \Psi^{m}_{j\mp \tfrac{1}{2}} \end{pmatrix}
\end{equation}
\begin{equation}
\Psi^{m}_{j-\tfrac{1}{2}}=\frac{1}{\sqrt{2j}}\begin{pmatrix}\sqrt{j+m}Y^{m-\tfrac{1}{2}}_{j-\tfrac{1}{2}}\\ \sqrt{j-m}Y^{m+\tfrac{1}{2}}_{j-\tfrac{1}{2}}\end{pmatrix}\text{\ and\ }\Psi^{m}_{j+\tfrac{1}{2}}=\frac{1}{\sqrt{2j+2}}\begin{pmatrix} \sqrt{j+1-m}Y^{m-\tfrac{1}{2}}_{j+\tfrac{1}{2}}\\ -\sqrt{j+1+m}Y^{m+\tfrac{1}{2}}_{j+\tfrac{1}{2}}\end{pmatrix}.
\end{equation}
\end{subequations}
We recall that the $Y^m_{\ell}$ are the spherical harmonics (eigenvectors of $\Lbf^2$).

Hence $E$ is spanned by a wave function which is a linear combination of that of type \eqref{di_irr_0}. 

We recall that for any integer $n\ge 1$ there is but one irreducible representation of $\mathbf{SU}(2)$ of dimension $n$ up to isomorphism. They can be found by the number of eigenvalues of $J_3'$, the infinitesimal "rotation" around the $z$ axis which induces a representation of $\mathbf{SO}(3)$.. Here $J'_3$ corresponds to $J_3$.

Thus we get that for $\eps\in\{+,-\}$
\[
E_{\eps}:=\PPh\,a(r)\Phi^\eps_{j,\kappa_j}
\]
is irreducible with respect to $\PPh$. By unicity of the irreducible representation of dimension $2j+1$, there exists an isomorphism from $E_-$ to $E_+$. As there must be a correspondence between the eigenspace of $J_3(E_-)$ and that of $J_3(E_+)$, necessarily $\mathbb{C}a\Phi^-_{m,\kappa_j}$ is sent to $\mathbb{C}a\Phi^+_{m,\kappa_j}$.

In particular as $\Pup E$ and $\Pdow E$ are also representation of $\mathbf{SU}(2)$ with same eigenvalues of $\Jbf^2,\Spbf$ (or $=\{0\}$). If one of them is zero then $E$ is of type $E_\pm$. If both are non-zero, then there exists $a_\uparrow(r),a_{\downarrow}(r)$ such that
\[
\Pup E=\PPh a_\uparrow(r)\Phi^{+}_{j,\kappa_j}\text{\ and\ }\Pdow E=\PPh a_{\downarrow}(r)\Phi^{-}_{j,\kappa_j}.
\]
Both $\Pup E$ and $\Pdow E$ are irreducible. We can suppose that there exists $f\in E$ with
\[
\Pup f= a_\uparrow(r)\Phi^{+}_{j,\kappa_j}\text{\ and\ }\Pdow f=a_{\downarrow}(r)\Phi^{-}_{j,\kappa_j}.
\]

The isomorphism between the two representations implies that
\[
E=\PPh\big( a_\uparrow(r)\Phi^{+}_{j,\kappa_j}+a_{\downarrow}(r)\Phi^{-}_{j,\kappa_j}\big).
\]


\subsection{Proof of Proposition \ref{di_mani_ci_sym}}

We have to prove that $\mathscr{M}_{\mathscr{I}}$ and $\mathscr{W}$ are submanifold of $\mathscr{M}$. The method is similar to the one used in \cite{pos_sok} to prove that $\mathscr{M}_{\mathscr{C}}$ is a submanifold of $\mathscr{M}$.

Let $P_0=Q_0+\PP\in \mathscr{M}$. We will prove that in a neighbourhood of $P_0$ in $\PP+\mathfrak{S}_2$, the projectors $P_1$ in $\mathscr{M}_{\mathscr{I}}$ (resp. $\mathscr{W}$) can be written as
\[
P_1=e^{A}P_0e^{-A},
\]
where $A\in\mathfrak{m}^{\mathscr{I}}_{P_0}$ (resp. $\mathfrak{m}^{\mathscr{W}}_{P_0}$). 

\noindent -- If we assume this point, then it is clear that the two sets are submanifolds of $\mathscr{M}$. Indeed $e^A$ is a global linear isometry of $\hl$, whose restriction to the $\mathfrak{m}_P^{\cdot}$'s maps $\mathfrak{m}_{P_0}^{\cdot}$ onto $\mathfrak{m}_{P_1}^{\cdot}$.

Equivalently it maps the first tangent plane onto the other:
\[
 \{ [a,P_0],\ a\in\mathfrak{m}_{P_0}^{\cdot}\}\underset{\simeq}{\to} \{ [a,P_1],\ a\in\mathfrak{m}_{P_1}^{\cdot}\}.
\]

\noindent -- We use Theorem \ref{di_structure} to write
\begin{equation}\label{di_di_chiant}
Q_0=\ssum_{j=1}^{+\infty}\la_j\big(\ket{f_j}\bra{f_j}-\ket{f_{-j}}\bra{f_{-j}}\big)
\end{equation}
where $(\la_i)_i\in\ell^2$ is non-increasing and the $f_i$'s form an orthonormal basis of $\mathrm{Ran}\,Q$. Provided that
\[
\ns{2}{P_1-P_0}<1,
\]
then $\la_1<1$ and there is no $j$ such that $f_j$ or $f_{-j}$ is in the range of $\PPP$ or $\PP$.

We decompose with respect with the eigenvalues $\mu_1>\mu_2>\cdots>0$ as follows:
\[
Q_0=\ssum_{k=1}^{+\infty}\mu_k\big(\text{Proj}\ \mathrm{Ker}(Q_0-\mu_k)-\text{Proj}\ \mathrm{Ker}(Q_0+\mu_k)\big).
\]
For short we write $\mu_{-k}:=-\mu_{k}$, and 
\begin{equation}M_k:=\text{Proj}\ \mathrm{Ker}(Q_0-\mu_k)\text{\ and\ }E_{\mu_k}^{Q_0}:=\mathrm{Ker}(Q_0-\mu_k).\end{equation}

As any $\YYY\in\{\Cha,\Isym\}$ is an isometry (linear or antilinear) and as the eigenvalues are the sine of the angles between vectors in $P_0$ and $\PP$,  for any $k$ we have
\begin{equation}\label{di_invar_y}
\YYY E_{\mu_k}^{Q_0}=E_{-\mu_k}^{Q_0}
\end{equation}
and the eigenspaces $E_{\mu_k}^{Q_0}\oplus E_{-\mu_k}^{Q_0}=\mathrm{Ker}(Q_0^2-\mu_k^2)$ are invariant under $\YYY$.

\paragraph{Case of $\mathscr{W}$}

\noindent -- In the case $\YYY=\Cha$ and $P_0\in \mathscr{W}$, each eigenspace 

\noindent $\mathrm{Ker}(Q_0^2-\mu_k^2)$ is also invariant under the action of $\PPh$. 
In other words, $\mathrm{Ker}(Q_0^2-\mu_k^2)$ is a finite dimensional representation of $\PPh$, and we can decompose it into irreducible representations $E_{\mu_k}^{(\ell)}$, where $0\le \ell\le \ell_k$.

By $\Cha$-symmetry, we have
\[
\Cha E_{\mu_k}^{(\ell_1)}=E_{-\mu_k}^{(\ell_1')},
\]
there is a one-to-one correspondence between irreducible representations of type $E_{\mu_k}^{(\ell)}$ and that of type $E_{-\mu_k}^{(\ell)}$. Up to changing indices $\ell'_j$, we can suppose that
\[
\Cha E_{\mu_k}^{(\ell)}=E_{-\mu_k}^{(\ell)},\ 0\le \ell\le \ell_k.
\]
Decomposing $E_{\mu_k}^{(\ell)}$ with respect with $\PP$ and $\PPP$, we see that
\[
\mathcal{P}^0_{\pm} E_{\mu_k}^{(\ell)}\ \text{is\ irreducible},
\]
and from the spectral decomposition of $Q_0$
\[
\PP E_{\mu_k}^{(\ell)}\oplus \PPP E_{\mu_k}^{(\ell)}=E_{\mu_k}^{(\ell)}\oplus F_{-\mu_k},
\]
where $F_{-\mu_k}$ is an irreducible subset of $\mathrm{Ker}(Q_0+\mu_k)$.

\noindent -- Let us show that
\begin{equation}\label{di_inters}
F_{-\mu_k}\cap \Cha E_{\mu_k}^{(\ell)}=\{ 0\}.
\end{equation}
Indeed, from Lemma \ref{di_irreduc} and the expression of the $\Phi^{\pm}_{m,\kappa}$, we see that
\[
\Cha \mathrm{Ker}\big(J_3-m\big)=\mathrm{Ker}\big(J_3+m\big).
\]
Thus if the intersection is non-zero, then we have by $\Cha$-symmetry and $\PPh$-symmetry:
\[
F_{-\mu_k}=\Cha E_{\mu_k}^{(\ell)}.
\]
But as shown in \cite{pos_sok}, this cannot happen: let us say that $E_{\mu_k}^{(\ell)}$ is associated to the eigenvalues $j_0(j_0+1),\kappa$ of $\Jbf^2$ resp. $\Spbf$. We consider:
\[
\mathrm{Ker}(J_3-m)\cap \mathcal{P}^0{\pm} E_{\mu_k}^{(\ell)}=\mathbb{C}e_{\pm;m},\ -j_0\le m\le j_0,\ \nlp{2}{e_{\pm;m}}=1.
\]
We would have
\[
\Cha e_{\pm;m}=\text{exp}{i\theta(\pm;m)}e_{\mp;-m}.
\]
The constant $\theta(\pm;m)$ does not depend on $m$ by $\PPh$-symmetry.
Moreover, if
\[
\mathrm{Ker}(J_3-m)\cap E_{\mu_k}^{(\ell)}=\mathbb{C}f_{m},
\]
then
\[
\mathcal{P}^0_{\pm} f_m\parallel e_{\pm;m}.
\]
As in \cite{pos_sok} for $\mathscr{M}_{\mathscr{C}}$, the condition $\Cha^2=1$ implies $\theta_+-\theta_-\equiv 0[2\pi]$ while
\[
-\Cha Q_0\Cha=Q_0
\]
implies $\theta_+-\theta_-\equiv \pi[2\pi]$, which cannot occur.

Similarly, we can prove that \eqref{di_inters} holds and that in fact $F_{-\mu_k}$ is orthogonal to $ \Cha E_{\mu_k}^{(\ell)}$.

As a consequence, the number of $E_{\mu_k}^{(\ell)}$'s is even, or equivalently, the number of $\PP E_{\mu_k}^{(\ell)}$ is even. 

\noindent -- The fact that
\begin{equation}\label{di_check}
P_1=e^A P_0 e^{-A},\ \text{with}\ \PPh A=A,\ \Cha A \Cha =A,\ \ns{2}{A}<+\infty,
\end{equation}
follows from Theorem \ref{di_structure} and the different symmetries.

The $f_j$'s in \eqref{di_di_chiant} can be written as ($\la_j=\sin(\theta_j)$)
\[
f_j=\sqrt{\frac{1-\la_j}{2}}e_{-;j}+\sqrt{\frac{1+\la_j}{2}}e_{+;j},\ \mathcal{P}^0_{\pm}e_{\pm;j}=e_{\pm;j}.
\]
We also have
\[
f_{-j}=-\sqrt{\frac{1+\la_j}{2}}e_{-;j}+\sqrt{\frac{1-\la_j}{2}}e_{+;j}.
\]
Then we define
\begin{equation}\label{di_aaa}
A=\ssum_{j=1}^{+\infty}\theta_j\big(\ket{e_{+;j}}\bra{e_{-;j}}-\ket{e_{-;j}}\bra{e_{+;j}}\big).
\end{equation}
It is easy to check that $A$ satisfies \eqref{di_check}. In fact, we can assume that $f_j$ spans an irreducible representation of $\mathbf{SU}(2)$, and in this case the same holds for $e_{+;j}$ and $e_{-;j}$.

As in Section \ref{di_irr_proof}, the correspondence $e_{-;j}\mapsto e_{+;j}$ induces an isomorphism between $\PPh e_{-;j}$ and $\PPh e_{+;j}$. This fact together with the $\PPh$-symmetry implies that
\[
 \forall\,U\in\mathrm{Ran}\,\PPh,\ UAU^{-1}=A.
\]
The fact that $\Cha A\Cha=A$ was proved in \cite{pos_sok} in the case $P_0,P_1\in\mathscr{M}_{\mathscr{C}}$. Here this remains true because
\[
 \mathscr{W}\subset \mathscr{M}_{\mathscr{C}}.
\]

\noindent --  We can now determine the connected component of $\mathscr{W}$. Let $P_0,P_1$ be in $\mathscr{W}$ and let $Q=P_1-P_0$.

We consider
\[
 E_1^{Q}:=\mathrm{Ker}(Q-1).
\]
If $E_1^Q=\{0\}$, then we can write $P_1=e^A P_0 e^{-A}$ as in \eqref{di_aaa}. And we see that the path in $\ell^2$:
\[
 t\in[0,1]\mapsto (t\theta_j)_j\in \ell^2
\]
induces a path connecting $P_0$ and $P_1$.

If $E_1^Q\neq \{ 0\}$, we count the number of irreducible representation in $E_1^Q$: let $b_{j,\kappa_j}$ be the number of irr. rep. in
\[
 \mathrm{Ker}\big(\Jbf^2-j(j+1)\big)\cap\mathrm{Ker}\big(\Spbf-\kappa_j\big).
\]
If all the $b_{j,\kappa_j}$'s are even, we can still write $P_1$ as $P_1=e^A P_0 e^{-A}$ with $A$ as in \eqref{di_aaa} with the first $\theta_j$ equal to $\tfrac{\pi}{2}$. In particular the two projectors can be connected by a path in $\mathscr{W}$.

Let us say that $b_{j_0,\kappa_{0}}\equiv 1[2]$ for some $j_0,\kappa_0$. We have shown that for $P\in\mathscr{W}$ with $\nb{P-P_0}<1$, the number of planes $\Pi_j$'s in the decomposition of Theorem \ref{di_structure} is even. Precisely, due to the $\Cha$-symmetry, there exists a sequence $(\ell_\mu(j,\kappa))_j$ in $\mathbb{N}$, with
\[
\begin{array}{l}
 \mathrm{Ker}\big((P-P_0)-\mu\big)\cap\mathrm{Ker}\big(\Jbf^2-j(j+1)\big)\cap \mathrm{Ker}\big(\Spbf-\kappa\big)\\
 \ \ \ =\underset{1\le \ell\le \ell_\mu(j,\kappa)}{\bigoplus}E^{(\ell)}_{\mu},
 \end{array}
\]
where each $E^{(\ell)}_{\mu}$ is irreducible as a representation of $\PPh$ and $\ell_\mu(j,\kappa)$ is \emph{even}.

We show that there cannot exist a continuous path linking $P_0$ and $P_1$ by a contradiction argument.

Let us say that $\g:t\in [0,1]\to \mathscr{W}$ is a continuous path with $\g(0)=P_0$ and $\nb{\g(1)-P_0}=1$. 

Then by the previous remarks, we have by continuity:
\[
\begin{array}{l}
\forall t\in[0,1], \forall\,j\in\frac{1}{2}+\mathbb{Z}_+,\ \forall\kappa\in\big\{\pm\big(j+\frac{1}{2}\big)\big\},\\
 \ \ \ \ \ell_1(Q_t=\g(t)-P_0;j,\kappa)\equiv 0[2].
 \end{array}
\]
In particular it is not possible to have $\g(1)=P_1$.


\paragraph{Case of $\mathscr{M}_{\mathscr{I}}$}

For $\YYY=\Isym$ and $P_0\in \mathscr{M}_{\mathscr{I}}$, we use \eqref{di_invar_y}. For each $f\in E_\mu^Q$, we have $\Isym \in E_{-\mu}^Q$ where $\mu\in \sigma(Q)$. We may assume that $\mu>0$.

Thus the plane
\[
 \Pi:=\text{Span}\big(f,\Isym f\big)
\]
is invariant under $Q$ and $\Isym$. We decompose $f$ and $\Isym f$ with respect to $P_0$ and $1-P_0$. By a dimension argument:
\begin{enumerate}
 \item either $\mu=1$, $P_0 f=0$ and $(1-P_0)\Isym f=0$,
 \item or $0<\mu<1$ and 
 \[
  \mathbb{C}P_0 f=\mathbb{C} P_0 \Isym f\text{\ and\ }\mathbb{C}(1-P_0) f=\mathbb{C} (1-P_0) \Isym f.
 \]
\end{enumerate}
In each case, we write $e_{-}$ a unitary vector in $\mathrm{Ran}\,P_0\cap \Pi$ and $e_+=\Isym e_-$. 

If we consider the sequence $(\mu_i)_i$ of positive eigenvalues of $Q$ (counted with multiplicities), we get the correspondent sequences $(e_{-;j})_j$ and $(e_{+;j})$. Moreover by Theorem \ref{di_structure}, we know that $\mu_j=\sin(\theta_j)$ where $\theta_j\in[0,\tfrac{\pi}{2}]$ is the angle between
the two lines $\mathbb{C}e_{-;j}$ and $\mathbb{C}f_j$. 

Provided that we take $-\theta_j$ instead of $\theta_j$ and up to a phase, we can suppose that
\[
 f_j=\cos(\theta_j)e_{-;j}+\sin(\theta_j)\Isym e_{-;j}.
\]

In particular we have
\[
 P_1=e^A P_0 e^{-A},
\]
with
\[
 A=\ssum \theta_j\big(\ket{e_{+;j}}\bra{e_{-;j}}-\ket{e_{-;j}}\bra{e_{+;j}}\big).
\]
It is straightforward to check that $\Isym A \Isym^{-1}=A$.






\medskip\medskip\medskip

\noindent\textit{Acknowledgment} The author wishes to thank \'Eric s\'er\'e for useful discussions and helpful comments. This work was partially supported by the Grant ANR-10-BLAN 0101 of the French Ministry of research.

\bibliographystyle{alpha}
{\small\bibliography{bibliothese.bib}}

\end{document}